\documentclass[11pt,a4paper]{article}
\usepackage{grffile}
\usepackage{aas_macros,braket}
\usepackage[T1]{fontenc} 
\usepackage{jcappub}
\usepackage[utf8]{inputenc}
\usepackage{amsmath}
\usepackage{amssymb}
\usepackage{bm}
\usepackage{url}
\usepackage{enumitem}
\usepackage[numbers,sort&compress]{natbib}
\usepackage[nottoc,notlot,notlof]{tocbibind}
\usepackage{wrapfig}
\usepackage[colorinlistoftodos]{todonotes}
\usepackage{color}
\usepackage{graphicx}
\usepackage{epsfig}
\usepackage{physics}
\usepackage[titletoc]{appendix}
\usepackage[normalem]{ulem}
\usepackage{framed}
\usepackage{tensor}
\usepackage{pdfcomment}
\usepackage{mathtools}
\usepackage{cancel}
\usepackage{geometry}
\usepackage{soul}
\usepackage{cleveref}
\crefrangeformat{equation}{(#3#1#4)-(#5#2#6)}

\newcommand{\ba}{\begin{align}}
\newcommand{\ea}{\end{align}}

\newcommand{\rhogw}{\ensuremath{\rho_{\rm GW}}}
\newcommand{\GW}{\ensuremath{{\rm GW}}}
\newcommand{\deltagw}{\ensuremath{\delta_{\rm GW}}}
\newcommand{\omegagw}{\ensuremath{\omega_{\rm GW}}}
\newcommand{\Omegagw}{\ensuremath{\Omega_{\rm GW}}}
\newcommand{\Omegabar}{\ensuremath{\overline{\Omega}_{\rm GW}}}
\newcommand{\tlgw}{\ensuremath{\widetilde{\Omega}_{\rm GW}}}
\newcommand{\etai}{\ensuremath{\eta_{\rm i}}}
\newcommand{\tl}[1]{\ensuremath{\Tilde{#1}}}
\newcommand{\hc}{\ensuremath{\mathcal{H}}}

\newcommand{\rc}{\ensuremath{\mathcal{R}}}
\newcommand{\pc}{\ensuremath{\mathcal{P}}}
\newcommand{\oc}{\ensuremath{\mathcal{O}}}

\newcommand{\vx}{\mathbf{x}}

\newcommand{\xmu}{x^{\mu}}
\newcommand{\pmu}{p^{\mu}}
\newcommand{\del}[2]{\frac{\partial #1}{\partial #2}}

\newcommand{\hn}{\hat n}
\newcommand{\mpbh}{M_{\rm PBH}}

\newcommand{\msun}{M_{\odot}}
\newcommand{\fpbh}{f_{\rm PBH}}

\newcommand{\SNR}{\text{SNR}}

\begin{document}
\title{Enhancing gravitational wave anisotropies with peaked scalar sources}
\author[a,b]{Ema Dimastrogiovanni,}\author[c,d]{Matteo Fasiello,}\author[b,e]{Ameek Malhotra,} \author[f]{Gianmassimo Tasinato}
\affiliation[a]{Van Swinderen Institute for Particle Physics and Gravity,
University of Groningen, Nijenborgh 4, 9747 AG Groningen, The Netherlands}
\affiliation[b]{School of Physics, The University of New South Wales, Sydney NSW 2052, Australia}
\affiliation[c]{Instituto de F\'{i}sica T\'{e}orica UAM/CSIC, Calle Nicol\'{a}s Cabrera 13-15, Cantoblanco, 28049, Madrid, Spain}
\affiliation[d]{Institute of Cosmology \& Gravitation, University of Portsmouth, PO1 3FX, UK}
\affiliation[e]{Sydney Consortium for Particle Physics and Cosmology, The University of New South Wales, Sydney NSW 2052, Australia}
\affiliation[f]{Physics Department, Swansea University, SA28PP, United Kingdom}

\emailAdd{e.dimastrogiovanni@rug.nl}
\emailAdd{matteo.fasiello@csic.es}
\emailAdd{ameek.malhotra@unsw.edu.au}
\emailAdd{g.tasinato@swansea.ac.uk}

\abstract{Gravitational wave (GW) backgrounds of cosmological origin are expected to be nearly isotropic, with small anisotropies resembling those of the cosmic microwave background. We analyse the case of a scalar-induced GW background and clarify in the process the relation between two different approaches to calculating GW anisotropies. We focus on GW scenarios sourced by a significantly peaked scalar spectrum, which are frequently considered in the context of primordial black holes production.~We show that the resulting GW anisotropies are characterised by a distinct frequency dependence. We explore the observational consequences concentrating on a GW background enhanced in the  frequency band of space-based GW detectors. We study  the detectability of the signal through both cross-correlations among different space-based GW detectors, and among GW and CMB experiments.
}

\maketitle
\flushbottom

\section{Introduction}
The detection of a stochastic gravitational wave background (SGWB) is one of the key targets of gravitational wave astronomy. 
Given the increasing sensitivity of ground based detectors \cite{LIGOScientific:2019vic,KAGRA:2021kbb}, and the tantalising hints  of a signal at NANOGrav \cite{NANOGrav:2020bcs}, the detection of a background of astrophysical origin is eagerly anticipated. In addition to the astrophysical background \cite{Regimbau:2011rp}, one should also expect the existence of a cosmological GW background (CGWB). Since the universe is, to a good approximation, transparent to gravitational waves below the Planck scale, the CGWB provides an excellent handle on the physics of the primordial universe (see for example the reviews in \cite{Caprini:2018mtu,Guzzetti:2016mkm}). 
\\
\indent The CGWB may arise from a variety of processes in the early universe, from inflation to later  sources including (p)re-heating dynamics, phase transitions, cosmic strings, and primordial black holes (PBH)  (see~\cite{Caprini:2018mtu} for a review). Remarkably, the GW signal associated to several such sources may be detected in the near future. Indeed, the corresponding frequencies are the one accessed by ground-based interferometers such as ET \cite{Punturo:2010zz}, CE \cite{Reitze:2019iox} $(1\text{--}10^3\,\rm Hz)$ and space-based experiments including LISA \cite{amaroseoane2017laser}, Taiji \cite{Hu:2017mde} $(10^{-4}\text{--}10^{-1}\,\rm Hz)$, all of which are expected to become operational by 2035.\\
\indent The GW background generated during inflation stands out in that (i) it is universal in nature and (ii) it may span a wide range of frequencies, from the CMB up to (and above) laser interferometer scales. In  single-field slow-roll  scenarios, the amplitude of this background is directly related to the energy scale at which inflation occurs\footnote{One should stress that this relation is not necessarily one-to-one. There 
are indeed very interesting cases that break this simple correspondence.
Such classes of models include so-called $P(X,\phi)$ theories (also known as k-essence), and the more general EFT of (single-field) inflation. One should also mention specific constructions with non-minimal couplings, such as those considered in \cite{Giare:2020plo} and \cite{Bartolo:2017szm}. Quite interesting in this context is also the setup known as ``Galileon inflation'' \cite{Burrage:2010cu},  which has received considerable attention. As to the origin of, for example, a subluminal sound speed in the (scalar) sector, one may point to the integrating out of additional fields in the inflationary Lagrangian, such as those kinetically coupled via the so-called gelaton mechanism of \cite{Tolley:2009fg}.}. Efforts to detect this background using the B-mode polarisation of the CMB have so far led to increasingly stringent constraints on the tensor-to-scalar ratio \cite{Ade:2015lrj,Akrami:2018odb,BICEP:2021xfz}, the most recent one being $r<0.032$ \cite{Tristram:2021tvh}. As a result of the ever-improving experimental bounds, various inflationary models have fallen by the wayside. Future CMB experiments like CMB-S4 \cite{CMB-S4:2020lpa}, Simons Observatory \cite{SimonsObservatory:2018koc} and LiteBIRD \cite{LiteBIRD:2022cnt} will be able to probe this background for values of the tensor-to-scalar ratio down to $r\sim 10^{-3}$ allowing us to further rule in(out) several classes of inflationary models.

The CMB-S4/LiteBIRD threshold on $r$ is particularly significant in that it will put to the test celebrated models of inflation such as the Starobinsky model \cite{Starobinsky:1980te} and (in the large field limit \cite{Kehagias:2013mya}) Higgs inflation. We stress therefore that, although direct detection of a SGWB from single-field slow-roll models is beyond the reach of most PTAs and interferometers (with the possible exception of the proposed BBO \cite{Crowder:2005nr} in the deciHz range), the lack of detection itself will be almost equally as informative. Crucially, GW detection by upcoming experiments would point to a number of interesting inflationary scenarios, including multi-field models (see e.g. \cite{Cook:2011hg,Barnaby:2011qe,Barnaby:2012xt,Dimastrogiovanni:2016fuu,Garcia-Bellido:2016dkw,Thorne:2017jft,Domcke:2018eki,Bordin:2018pca,Iacconi:2019vgc,Campeti:2020xwn,Iacconi:2020yxn}), theories featuring alternative symmetry breaking patterns \cite{Endlich:2012pz,Bartolo:2015qvr,Ricciardone:2016lym,Celoria:2020diz,Celoria:2021cxq}, and non-attractor phases \cite{Mylova:2018yap,Ozsoy:2019slf} (see \cite{Guzzetti:2016mkm} for a comprehensive review).\vspace{0.2cm} \\
\indent A SGWB detection will also provide the opportunity to test its dependence on direction. Most cosmological sources of GWs are known to be predominantly isotropic, with a small level of anisotropy that may arise from i) the specific production mechanism and ii) propagation in a perturbed universe. While the former anisotropies are typically model-dependent\footnote{For example, anisotropies can be induced by primordial squeezed non-Gaussianity in the tensor sector \cite{Dimastrogiovanni:2019bfl,Adshead:2020bji,Malhotra:2020ket,Dimastrogiovanni:2021mfs,Dimastrogiovanni:2022afr}.} the latter are fairly universal in nature, being a consequence of the fact that the universe is not perfectly homogeneous and isotropic \cite{Alba:2015cms,Contaldi:2016koz,Bartolo:2019oiq,Bartolo:2019yeu,Domcke:2020xmn}. The anisotropies of the CGWB may therefore hold  precious information on the early universe\footnote{SGWB anisotropies have also been studied in the context of phase transitions \cite{Geller:2018mwu,Kumar:2021ffi,Li:2021iva}, cosmic strings \cite{Olmez:2011cg,Kuroyanagi:2016ugi,Jenkins:2018nty,Cai:2021dgx}, as probes of $\Delta N_{\rm eff}$ \cite{DallArmi:2020dar}, and several pre-recombination scenarios \cite{Braglia:2021fxn}. A detailed analysis of their cross-correlation with the CMB temperature and E-mode polarisation can be found in \cite{Ricciardone:2021kel}.}. Understanding their behaviour is also a critical step towards being able to tell them apart from anisotropies of astrophysical origin \cite{Cusin:2017fwz,Cusin:2018rsq,Jenkins:2018uac,Jenkins:2018kxc,Jenkins:2019uzp,Cusin:2019jhg,Cusin:2019jpv,Bertacca:2019fnt,Pitrou:2019rjz,Capurri:2021zli,Bellomo:2021mer}, which act as a foreground for the CGWB.\\ 
\indent Propagation anisotropies \cite{Alba:2015cms} are similar in origin to those of the CMB and have been studied via a Boltzmann approach in~\cite{Contaldi:2016koz,Bartolo:2019oiq,Bartolo:2019yeu}. We will point out how the two approaches (in \cite{Alba:2015cms} and \cite{Contaldi:2016koz,Bartolo:2019oiq,Bartolo:2019yeu} respectively) are related to each other under certain assumptions about the behaviour of the primordial perturbations as well as the time when GW are generated.\vspace{0.2cm}\\ 
\indent  An important, ever-present, source of cosmological GW is the background sourced at second order in the primordial curvature perturbations \cite{Matarrese:1997ay,Ananda:2006af,Baumann:2007zm,Domenech:2021ztg}. At linear order in perturbation theory tensor and scalar modes are decoupled from each other. However, this is no longer the case at second order: as the primordial scalar perturbations re-enter the horizon, they source GW via their second order anisotropic stress. Whenever these scalar perturbations have spectra peaked at scales which re-enter the horizon during the radiation domination epoch, the resulting scalar induced gravitational waves (SIGW) may have an amplitude large enough to be detected at interferometer scales. 
For sufficiently peaked spectra, possible in inflationary models with significant primordial black hole (PBH) production (see e.g. \cite{Sasaki:2018dmp} for a review), this small-scale enhancement imparts a distinct spectral shape to the SIGW and its anisotropies. We will show how this also leads to a significant enhancement in the anisotropy spectrum at certain frequencies, with potentially observable consequences.  The frequency dependence of the anisotropies is present not just for the SIGW, but also for any other SGWB whose spectral shape deviates from a power law (e.g. (p)reheating, phase transitions, cosmic strings \cite{Caprini:2018mtu}). See also \cite{Auclair:2022lcg} for a general review.  \\

This paper is organised as follows. In Sec.~\ref{sec:SGWB_anisotropies} we first review the Boltzmann approach to the SGWB anisotropies and discuss the role of the GW initial conditions. We then demonstrate how and under what assumptions the result of~\cite{Alba:2015cms} can be related to those of the Boltzmann approach. In Sec.~\ref{sec:Anisotropy_PBH} we discuss the frequency dependence of the anisotropy spectrum for the SIGW. We show that if the primordial scalar power spectrum is sharply peaked at certain scales, the resulting SIGW anisotropies can be enhanced 
at certain frequencies by a factor $\oc(10\text{--}100)$ relative to the anisotropies for a standard power-law spectrum. We then discuss the implications of this enhancement for the detection of GW anisotropies using LISA-Taiji and BBO. Finally, we present our conclusions in Sec.~\ref{sec:conclusion}.

\section{Anisotropies of SGWB}
\label{sec:SGWB_anisotropies}
In this section we start by reviewing the propagation anisotropies of the SGWB, both via the Boltzmann approach of refs.~\cite{Contaldi:2016koz,Bartolo:2019oiq,Bartolo:2019yeu} and via the calculation in \cite{Alba:2015cms}. As discussed in ref.~\cite{Contaldi:2016koz}, the Boltzmann formalism relies on the geometrical optics approach to GW propagation which is valid for GW whose wavelengths are much smaller than the length scales over which the background curvature varies \cite{Isaacson_geomoptics,Misner:1973prb}. The same holds for the approach of \cite{Alba:2015cms}. Later in this section we discuss the relation between these results and the role of the GW initial conditions. The results we develop here will be applied in Sec.~\ref{sec:Anisotropy_PBH} to study  GW anisotropies from scalar induced gravitational waves.

\subsection{Boltzmann approach to SGWB anisotropies}
\label{sec:Boltzmann_NG}
We begin with a brief review of the main results of \cite{Contaldi:2016koz,Bartolo:2019oiq,Bartolo:2019yeu}. The starting point is the distribution function for gravitons $f(\xmu,\pmu)$\footnote{Note that, as a consequence of equivalence principle, the stress-energy carried by gravitational waves cannot be localised within their wavelength. Thus, the energy density should be understood as obtained by averaging over length(time)-scales much larger than the wavelength(time)-scales associated to the GW \cite{Isaacson_geomoptics,Misner:1973prb}.}. 
Here $\xmu,\pmu$ denote the position and momentum of the gravitons with $\pmu=d\xmu/d\lambda$, $\lambda$ being the affine parameter along the graviton trajectory which, in the geometric optics limit, is given by the null geodesics of the perturbed spacetime. The evolution of the distribution function is governed by the Liouville equation  
\begin{align}
        \frac{df}{d\lambda} = C[f(\lambda)] + I[f(\lambda)],
\end{align}
where $C$ and $I$ are the collision and injection terms. The collision term is generally absent since gravitons are decoupled below the Planck scale. For stochastic backgrounds of cosmological origin, the injection term can be treated as an initial condition of the distribution \cite{Bartolo:2019oiq,Bartolo:2019yeu}. Thus, one must solve the free Boltzmann equation $df/d\lambda= 0$ in the perturbed universe. One can write it as 
\begin{align}
\label{eq:Boltzmann_eq}
 \frac{df}{d\eta} = \del{f}{\eta} + \del{f}{x^i}\frac{dx^i}{d\eta} + \del{f}{q}\frac{dq}{d\eta}=0.   
\end{align}
Here, 
\begin{equation}
q \,\equiv\,|\vec{p}|a
\end{equation}
is the comoving momentum of the gravitons, $a$ is the cosmological scale factor, and $\hn=\hat p$ is the direction of propagation. In the Newtonian gauge with $\Phi$ and $\Psi$ as the large-scale scalar potentials
we have,
\begin{align}
    \label{eq:metric_NG}
    ds^2 = a^2(\eta)\left[-(1+2\Phi)d\eta^2 + (1-2\Psi)\delta_{ij}dx^i dx^j\right],
\end{align}
and the Boltzmann equation at first order in $\Phi,\Psi$ can be written as 
\cite{Bartolo:2019oiq,Bartolo:2019yeu},
\begin{align}
\label{one-eq}
    \del{f}{\eta} + \del{f}{x^i}n^i+ q\del{f}{q}\left[\del{\Psi}{\eta}-\del{\Phi}{x^i}n^i\right]  =0.
\end{align}
 The total SGWB energy density observed today is given by  
\begin{align}
    \rhogw(\eta_0,\vec{x}_0) = \int d^3p\,p f(\eta_0,\vec{x_0},q,\hn)\,,
\end{align} and the more commonly used fractional density parameter $\Omegagw(q)$ is defined as \cite{Maggiore:1999vm}
\begin{align}
    \rhogw(\eta_0,\vec{x}_0) = \rho_{cr}\int d\ln q\, \Omegagw(\eta_0,\vec{x}_0,q), 
\end{align}
where $\rho_{cr}$ is the critical energy density of the universe.
Upon expanding the distribution function into a homogeneous and isotropic part $\bar{f}(q)$ and a perturbation $\delta f = -q\Gamma(\eta,\vec{x},q,\hn)\partial{\bar{f}}/\partial{q}$ one obtains the following (very reminiscent of the analogous one in the CMB context) equation \cite{Bartolo:2019oiq,Bartolo:2019yeu},
\begin{align}
        \label{eq:Gamma_NG}
        \Gamma(\eta_0,\vec{x}_0,\hn,q) = \underbrace{\Gamma(\etai,\vec{x}_i,q)}_{\Gamma_I} + \underbrace{\Phi(\etai,\vec{x}_i) + \int_{\etai}^{\eta_0} d\eta (\Phi'+\Psi')}_{\Gamma_S}, 
\end{align}
where $\eta_0,\vec{x}_0$ correspond to the time and position of observation and one has $\vec x_i = \vec x_0 - (\eta_0-\eta_i)\hn$. The first term on the right-hand side (R.H.S). denotes the initial perturbation at the time of emission $\eta_i$ while the second and third terms arise due to propagation in an inhomogeneous universe, analogous to the Sachs-Wolfe (SW) and integrated Sachs-Wolfe (ISW) effects for the CMB. As pointed out in \cite{Bartolo:2019yeu}, for adiabatic initial conditions the term $\Gamma_I$ is also correlated with the scalar term and contributes to the SW effect for gravitons. 
We explicitly evaluate this term in Sec.~\ref{sec:ad_ic} for the case of single clock inflation. \\

\indent The GW anisotropy observed today is defined as \cite{Bartolo:2019yeu,Bartolo:2019oiq},
\begin{align}
    \label{eq:deltagw_Boltzmann}
    \deltagw(\eta_0,\vec{x}_0,\hn,q) \equiv \frac{\omegagw(\eta_0,\vec{x}_0,\hn,q)}{\Omegabar(\eta_0,q)}-1 =  \left[4 - \del{\ln \Omegabar(\eta_0,q)}{\ln q}\right] \Gamma(\eta_0,\vec{x}_0,\hn,q),
\end{align}
where the quantity $\omegagw$ is given by
\begin{align}
    \Omegagw(\eta,\vec{x},q) = \frac{1}{4\pi}\int d^2\hn\, \omegagw(\eta,\vec{x},q,\hn),
\end{align}
with $\Omegabar$ denoting the spatial average of $\Omegagw(q,\vec{x})$. In terms of the distribution function we have \cite{Bartolo:2019yeu,Bartolo:2019oiq},
\begin{align}
\label{eq:omegabar_f}
    \Omegabar(q,\eta) = \frac{4\pi}{\rho_{\rm cr}}\left(\frac{q}{a}\right)^4\,\bar{f}(q).
\end{align}

\noindent We can now calculate how the initial condition $\Gamma_I$ is related to the initial density fluctuation $\delta\rhogw$, with $\rhogw(\eta_i,\vx)\equiv\bar{\rho}_\GW(\eta_i)+\delta\rhogw(\eta_i,\vx)$. This will then be used in Sec.~\ref{sec:ad_ic} to obtain $\Gamma_I$ in terms of the primordial potential $\Phi$. One finds:
\begin{align}
    \rhogw(\eta_i,\vec{x}) &= \int d^3p\, p\bar{f}(q)\left[1-\del{\ln \bar{f}}{\ln q}\Gamma(\eta_i,\vec{x},q,\hn)\right] \nonumber \\
    &=\frac{4\pi}{a^4}\int dq\, \bar{f}(q)q^3 - \frac{1}{a^4}\int dq \frac{\partial \bar{f}}{\partial q}q^4\int d^2\hn\,\Gamma_I,\quad \text{with }\Gamma_I\equiv\Gamma(\eta_i,\vec{x},q,\hn)\,.
\end{align}
If the only contribution to $\Gamma_I$ arises from the adiabatic primordial perturbations, which is the scenario we are interested in here, $\Gamma_I$ can be safely assumed to be $q$-independent\footnote{{In general $\Gamma_I$ can indeed be $q$-dependent, the specific form of this dependence arises from the production mechanism of the SGWB. An example of this $q$-dependence is discussed in ref.~\cite{Bartolo:2019yeu}. In case such a $q$-dependence is present, the correlators of the observed anisotropy $\Gamma$ are no longer frequency dependent and one has $ \Gamma(q,\hat n)=\sum_{\ell m}\Gamma_{\ell m}(q)Y_{\ell m}(\hat n)$.}}. In this case, upon integrating by parts the integral over $q$ on the R.H.S., one arrives at 
\begin{align}
    \rhogw(\eta_i,\vec{x})= \frac{4\pi}{a^4}\int dq\, \bar{f}(q)q^3\left(1+ 4\int \frac{d^2\hn}{4\pi}\,\Gamma_I\right) \equiv \bar{\rho}_{\rm GW} + \delta\rhogw.
\end{align}
From here one can see that the intrinsic density perturbation in the Newtonian gauge at the initial time $\eta_i$ is related to the monopole of $\Gamma_I$ by
\begin{align}
\label{eq:gammaI_rho}
    \frac{\delta\rhogw}{\bar\rho_{\rm GW}} = 4\int \frac{d^2\hn}{4\pi}\,\Gamma_I \equiv 4\Gamma_I^{(0)}.
\end{align}
We are already familiar with this result from the CMB ($\Gamma = \Theta \equiv \delta T/T$), where we have $\delta\rho_\gamma/\bar{\rho}_\gamma\equiv\delta_{\gamma} = 4\Theta_0$, $\Theta_0$ being the intrinsic temperature fluctuation at recombination. Note that in our analysis we will neglect any higher order multipole terms in $\Gamma_I$ and therefore from now on we will simply identify $\Gamma_I=\Gamma_I^{(0)}=\delta\rho_\GW/(4\bar{\rho}_\GW)$.

\vspace{0.2cm}
\noindent \textbf{When to evaluate initial conditions.}
For GW detectable at interferometer scales, the initial conditions are set early during radiation domination (RD) since that is when these GW are generated (e.g. from 1st order phase transitions, second order GW from scalar perturbations). Similarly, for inflationary GW to be tested e.g. at interferometers, RD is when the GW modes re-enter the horizon and start to propagate freely. 
It is only once the GW modes become sub-horizon ($q\ll \hc$ ) that $\rhogw\propto a^{-4}$ and one can correspondingly think of them as behaving like relativistic particles. Instead, for superhorizon modes ($q\gg \hc$) the energy density scales as $\rhogw \propto a^{-2}$ \cite{Abramo:1997hu}.

\subsubsection{Initial conditions for GW anisotropies}
\label{sec:ad_ic}
We now calculate the initial GW overdensity $\delta\rhogw(\eta_i,\vx)$ at the time of emission $\eta_i$, assuming that the primordial perturbations from inflation are adiabatic. These perturbations correspond to common, local time shifts in all background quantities as a result of which they have the property that for two different species $i$ and $j$,
\begin{align}
    \frac{\delta\rho_i}{(1+\mathsf{w}_i)\bar{\rho}_i}=\frac{\delta\rho_j}{(1+\mathsf{w}_j)\bar{\rho}_j}\,,
\end{align}
where $\mathsf{w}_i$ denotes the equation of state parameter for species $i$. Keeping this in mind, we can now easily relate the initial GW density perturbation to the potential $\Phi$ (we neglect any anisotropic stresses so that $\Phi=\Psi$). Firstly, note that the super-horizon solution for the photon  density contrast during radiation domination is given by \cite{Dodelson:2003ft},
\begin{align}
    \delta_\gamma = -2\Phi\,.
\end{align}
Therefore, by adiabaticity (neglecting sub-leading slow-roll corrections) and using the fact that $\mathsf{w}_\GW=\mathsf{w}_\gamma=1/3$, we have\footnote{We should stress here that, in our notation, ${\delta\rhogw}/{\bar{\rho}_{\rm GW}}$ is not the same as the quantity $\deltagw$ defined in Eq.~\eqref{eq:deltagw_Boltzmann}. $\delta\rhogw$ denotes the perturbation to the GW energy density $\bar{\rho}_\GW$ whereas $\deltagw$ is the anisotropy in the fractional energy density parameter $\Omegabar$.},
\begin{align}
    \label{eq:ad_ic_phi}
    \frac{\delta\rhogw}{\bar{\rho}_{\rm GW}} = \delta_\gamma = -2\Phi\ \implies \Gamma_I = \frac{1}{4}\frac{\delta\rhogw}{\bar{\rho}_{\rm GW}} = -\frac{1}{2}\Phi\,,
\end{align}
with all the quantities evaluated at the initial time $\eta_i$.

\vspace{0.2cm}
\noindent Alternatively, one can obtain the result of Eq.~\eqref{eq:ad_ic_phi} for the initial density perturbation in the Newtonian gauge by first converting the metric element to an unperturbed form 
\begin{align}
   ds^2 = a^2(\tilde{\eta})\left[-d\tilde{\eta}^2 + \delta_{ij}d\tilde{x}^i d\tilde{x}^j\right]
\end{align}
via the coordinate transformation $x^\mu\to \tl{x}^{\mu}\equiv x^\mu + \xi^\mu$ with \cite{Creminelli:2011sq}
\begin{align}
    \label{eq:transform_FLRW}
    \xi^0 = 
    \begin{dcases}
      -\frac{1}{5}\zeta\eta, & \text{M.D.} \\
      -\frac{1}{3}\zeta\eta, & \text{R.D.}
     \end{dcases}
\end{align}
and
\begin{align}
        \xi^i & = -\zeta x^i\,.
\end{align}
In the unperturbed coordinates we have $\delta\tilde{\rho}_\GW=0$. Thus, the density perturbation $\delta\rhogw$ in the Newtonian gauge is, 
\begin{align}
    \delta\rhogw & = \delta\tilde{\rho}_\GW +\bar{\rho}_\GW'\xi^0 \nonumber \\
    &= -4\hc\bar{\rho}_\GW\xi^0\,,
\end{align}
where $\bar{\rho}_\GW' = -3\hc(1+\mathsf{w}_\GW)\bar{\rho}_\GW$. During radiation domination one then finds, 
\begin{align}
    \frac{\delta\rhogw}{\bar{\rho}_{\rm GW}} = -2\Phi\,,
\end{align}
where we used the radiation domination expression for $\xi^0$ given in Eq.~\eqref{eq:transform_FLRW} and the relations $a\propto \eta$, and $\Phi= -2\zeta/3$. In contrast, for the CMB the relevant epoch for the initial conditions is that of matter domination and one obtains the standard result \cite{Creminelli:2011sq}
\begin{align}
    \label{eq:MD_deltarho}
    \delta_\gamma = -4\hc\xi^0 = -\frac{8}{3}\Phi,
\end{align}
where the matter domination relations $a\propto \eta^2$ and $\Phi = -3\zeta/5$ have been used.\\

\noindent Putting together Eqs.~\eqref{eq:Gamma_NG} and \eqref{eq:ad_ic_phi}, the total anisotropy observed today for GW produced during the radiation dominated epoch, under the assumption of adiabatic initial conditions, is given by
\begin{align}
    \label{eq:GW_ad}
    \Gamma(\eta_0,\vec{x}_0,\hn) &= \underbrace{\Gamma(\eta_i,\vec{x}_i)}_{\Gamma_I} + \Phi(\eta_i,\vec{x}_i)+\int_{\etai}^{\eta_0} d\eta (\Phi'+\Psi') \nonumber \\
    & = \frac{1}{2}\Phi  (\eta_i,\vec{x}_i)+\int_{\etai}^{\eta_0} d\eta (\Phi'+\Psi') \\
    & = -\frac{1}{3}\zeta(\eta_i,\vec{x}_i)+\int_{\etai}^{\eta_0} d\eta (\Phi'+\Psi')\,, \nonumber
\end{align}
where we remind the reader that $\vec{x}_0 = \vec{x}_i + (\eta_0-\eta_i)\hn$. Thus, although the large scale GW anisotropies have the same origin as those of the CMB, the coefficient appearing in front of the potential $\Phi$ is $1/2$ instead of $1/3$ due to the fact that the gravitons of interest (for direct detection at intermediate and small scales) start propagating during radiation domination. Indeed, if instead recombination took place during radiation domination, then the CMB SW effect would also be modified accordingly \cite{Hwang:2001hr}.\\

The effect of primordial isocurvature perturbations on the GW anisotropy has also been studied recently in \cite{Kumar:2021ffi,Bodas:2022zca}. The isocurvature perturbations modify the GW anisotropy as  \cite{Kumar:2021ffi},
\begin{align}
     \Gamma_{\rm iso} = \Gamma_{\rm ad} + \frac{1}{3}(1-f_\GW)S_\GW\,.
\end{align}
Here $f_\GW\equiv \rho_{\text{GW}}/\rho_{\text{tot}}$, with $\rho_{\text{tot}}$ the total energy density, and $\Gamma_{\rm ad}$ is the GW anisotropy, assuming adiabatic initial conditions, given in Eq.~\eqref{eq:GW_ad}. $S_\GW$ is the GW isocurvature perturbation w.r.t photons defined as $S_{\GW} \equiv 3(\zeta_{\GW}-\zeta_\gamma)$, with $\zeta_i = -\Psi - H{\delta\rho_i}/{\dot{\bar{\rho}}_i}$ the curvature perturbation on the hyper-surfaces of uniform energy of the fluid $i$.

\subsection{Uniform density gauge calculation}
\label{sec:Boltzmann_zeta}
We now compute the propagation anisotropy following ref.~\cite{Alba:2015cms}. The starting point for the calculation of \cite{Alba:2015cms} is the uniform density gauge during matter domination for which the metric is given by \cite{Boubekeur:2008kn},
\begin{align}
\label{eq:zeta_gauge}
    ds^2 = a^2(\eta)\left[-d\eta^2 + (1+2\zeta)\delta_{ij}dx^i dx^j -\frac{4}{5aH}\partial_i\zeta d\eta dx^i\right]\,.
\end{align}
To facilitate the comparison with the results of the previous section, we compute here the GW anisotropy via the Boltzmann approach working in the uniform density gauge given by Eq.~\eqref{eq:zeta_gauge}. Although ref.~\cite{Alba:2015cms} proceeds in a slightly different manner, the results are equivalent. For completeness we also report the original calculation of ref.~\cite{Alba:2015cms} as well as some details related to the Boltzmann approach in Appendix~\ref{app:Boltzmann_zeta}. 

To solve the Boltzmann equation \eqref{eq:Boltzmann_eq}, we express more conveniently its various terms. We write the graviton 4-momentum $P^\mu$ in terms of the magnitude of the physical momentum $p$ and the direction of propagation $\hn$ as (see Appendix~\ref{app:Boltzmann_zeta} for the details)
\begin{align}
    P^\mu = \frac{p}{a}\left(1-\frac{2}{5aH}\partial_i\zeta n^i, (1-\zeta)\hn\right)\,.
\end{align}
From this one can simply read the derivative  
\begin{align}
    \frac{dx^i}{d\eta} = \frac{P^i}{P^0} = n^i\,.
\end{align}
Note that we only need this term at zeroth order, since it multiplies $\partial f/\partial x^i$ which is a first order quantity. Next, we need $dq/d\eta$. After some manipulations (see Appendix~\ref{app:Boltzmann_zeta}), one finds
\begin{align}
    \frac{dq}{d\eta} = \frac{1}{5}q\,\partial_i\zeta n^i\,.
\end{align}
Using the above results, one can finally write 
\begin{align}
    \label{eq:Boltzmann_zeta_2}
    \del{f}{\eta} + \del{f}{x^i}n^i + \frac{1}{5}q\partial_i\zeta n^i \del{f}{q} = 0\,.
\end{align}
In terms of the zeroth order distribution $\bar{f}$ and the first order perturbation $\delta f = -q\Gamma(\eta,\vec{x},q,\hn)\partial{\bar{f}}/\partial{q}$ one obtains,
\begin{align}
    \del{\Gamma}{\eta} + n^i\del{\Gamma}{x^i} &= \frac{1}{5}\del{\zeta}{x^i}n^i\,.
\end{align}
This equation can be integrated along the line of sight to give,
\begin{align}
     \label{eq:Gamma_zeta}
     \Gamma(\eta,\vec{x}_0,q,\hn) = \underbrace{\Gamma(\eta_{\rm in},\vec{x}_i,q,\hn)}_{\Gamma_I} -\underbrace{\frac{1}{5}\zeta(\eta_{\rm in},x_i)}_{\Gamma_S}\,,      
\end{align}
where $\vec{x}_0 = \vec{x}_i + (\eta-\eta_{\rm in})\hn$. The curvature perturbation at the observer's position can be ignored since it does not have any direction dependence.   
Note that the Authors of ref.~\cite{Alba:2015cms} consider the case of single-field inflation and assume adiabatic primordial perturbations. Since the gauge of Eq.~\eqref{eq:zeta_gauge} is the uniform matter density gauge, it must also be the uniform GW density gauge, i.e. by adiabaticity one finds
\begin{align}
    \frac{\delta\rhogw}{\bar{\rho}_\GW}=\frac{4\delta\rho_{\rm m}}{3\bar{\rho}_{\rm m}}=0\implies \Gamma_I = 0,
\end{align}
where we have used Eq.~\eqref{eq:gammaI_rho}. Using this result in Eq.~(\ref{eq:Gamma_zeta}), the relation in Eq.~\eqref{eq:deltagw_Boltzmann} between the GW anisotropy $\deltagw$ and $\Gamma$, and Eq.~\eqref{eq:omegabar_f}, one obtains
\begin{align}
    \label{eq:deltagw_Alba}
    \deltagw(q,\eta_0,\hn) &= -\frac{1}{5}\zeta(\eta_i,\vec{x}_i)\left[4-\del{\ln\Omegabar}{\ln q}\right]=\frac{1}{5}\zeta(\eta_i,\vec{x}_i)\del{\ln \bar{f}(q)}{\ln q}\,,
\end{align}
which is the result of ref.~\cite{Alba:2015cms}. Let us now see how to relate this to the Newtonian gauge result of Eq.~\eqref{eq:GW_ad}. Firstly, note that under the assumption of matter domination one can simply disregard the ISW term from that equation and write
\begin{align}
    \label{eq:NG_comparison}
    \Gamma = \frac{1}{4}\frac{\delta\rhogw}{\bar{\rho}_\GW}+\Phi\,.
\end{align}
In this equation, we now use the matter domination relation, Eq.~\eqref{eq:MD_deltarho}, together with the adiabaticity condition, to get $\delta_\gamma = \delta\rhogw/\bar{\rho}_\GW = -8\Phi/3$. Thus Eq.~\eqref{eq:NG_comparison} becomes,
\begin{align}
    \Gamma = \frac{1}{3}\Phi = -\frac{1}{5}\zeta 
\end{align}
and
\begin{align}
    \deltagw &= -\left[4-\del{\ln\Omegabar}{\ln q}\right]\frac{\zeta}{5}\,,
\end{align}
which matches Eq.~\eqref{eq:deltagw_Alba}. However, we stress once again that given the choice of the metric in Eq.~\eqref{eq:zeta_gauge}, the result of ref.~\cite{Alba:2015cms} is valid only if we assume matter domination throughout. As mentioned in Sec.~\ref{sec:Boltzmann_NG}, for most cosmological SGWB the initial conditions are instead evaluated during radiation domination and the corresponding expression is that of Eq.~\eqref{eq:GW_ad}.

\section{Anisotropies for peaked spectra}
\label{sec:Anisotropy_PBH}

\noindent We now use the results of Sec.~\ref{sec:SGWB_anisotropies} to calculate  the propagation anisotropy spectrum associated to a cosmological background of gravitational waves sourced at second order in the curvature perturbations which re-enter the horizon during the epoch of radiation domination \cite{Ananda:2006af,Baumann:2007zm}. We are interested in scenarios where the primordial curvature power spectrum is sharply peaked on small scales. We parametrize these primordial spectra around the peak wavenumber $k_*$ by a log-normal, i.e. we take 
\begin{align}
    \label{eq:lgnormal_pk}
    \mathcal{P_{\rc}}(k)|_{k\gg k_{\rm CMB}} = \frac{A_{\mathcal{R}}}{\sqrt{2\pi}\Delta}\exp\left[-\frac{\ln^2(k/k_*) }{2\Delta^2}\right]\,.
\end{align}
The parameter $\Delta$ then controls the width of the spectrum and $A_{\rc}$ represents its amplitude. In the limit $\Delta \to 0$ one recovers the Dirac-delta power spectrum $\pc_{\rc} = A_{\rc} \delta (\ln(k/k_*))$. The log-normal serves as a useful representative of a peaked spectrum which can arise in several inflationary models producing PBH (see \cite{Pi:2020otn} and references therein). Note that the quantity $\rc$ is the comoving gauge curvature perturbation which on super-Hubble scales is equivalent to the uniform density gauge curvature perturbation $\zeta$.
\subsection{Scalar induced GW background and its anisotropies}
The GW energy density spectrum observed today for a log-normal power spectrum was calculated in \cite{Pi:2020otn} and can be written as
\begin{align}
\label{eq:SIGW_expr}
    \Omegagw(k,\eta_0)h^2\simeq 1.6\times10^{-5}\left(\frac{g_{*s}(\eta_k)}{106.75}\right)^{-1/3}\left(\frac{\Omega_{r,0}h^2}{4.1\times 10^{-5}}\right)\Omega_{\rm GW,r}(k)\,,
\end{align}
where $\Omega_{\rm GW,r}$ is the GW energy density at matter radiation equality and is given by 
\begin{align}
    \Omega_{\rm GW,r}(k) = 3\int_0^\infty dv\int_{|1-v|}^{1+v}du \frac{\mathcal{T}(u,v)}{u^2 v^2}\mathcal{P}_{\mathcal{R}}(vk)\mathcal{P}_{\mathcal{R}}(uk)\,,
\end{align}
with the function $\mathcal{T}(u,v)$ defined in Eq.(10) of \cite{Pi:2020otn}. A generic GW background of cosmological origin will have anisotropies in the angular distribution of its energy density. These anisotropies\footnote{In this section we focus only on the propagation anisotropies of the SGWB and therefore the quantity $C_{\ell}^\Gamma$ is frequency independent \cite{Bartolo:2019yeu}. We neglect any other sources of anisotropies that might be relevant for scalar induced GW, e.g. those arising from primordial non-Gaussianity due to a local $\langle \zeta^3 \rangle$ bispectrum \cite{Bartolo:2019zvb}.} are given by 
\begin{align}
    \label{eq:def_deltagw}
    \deltagw(k,\hn) =\left(4-n_{\Omega}(k)\right)\Gamma(k,\hn),\quad n_{\Omega}(k) \equiv {\partial \ln \Omegabar}/{\partial \ln k}\,,
\end{align}
where $\Gamma$ was defined in Eq.~\eqref{eq:Gamma_NG}. Thus, if the spectrum is sharply peaked ($|n_{\Omega}(k)|\gg 1$) one can expect enhancement of $C_\ell^\GW$ relative to $C_{\ell}^\Gamma$ (see Eqs. \ref{eq311}, \ref{eq312}) by a factor $\sim \mathcal{O}(10 \text{--}1000)$ at certain scales.  
\begin{figure}
\centering
\begin{tabular}{c c}
    \includegraphics[width=0.45\linewidth]{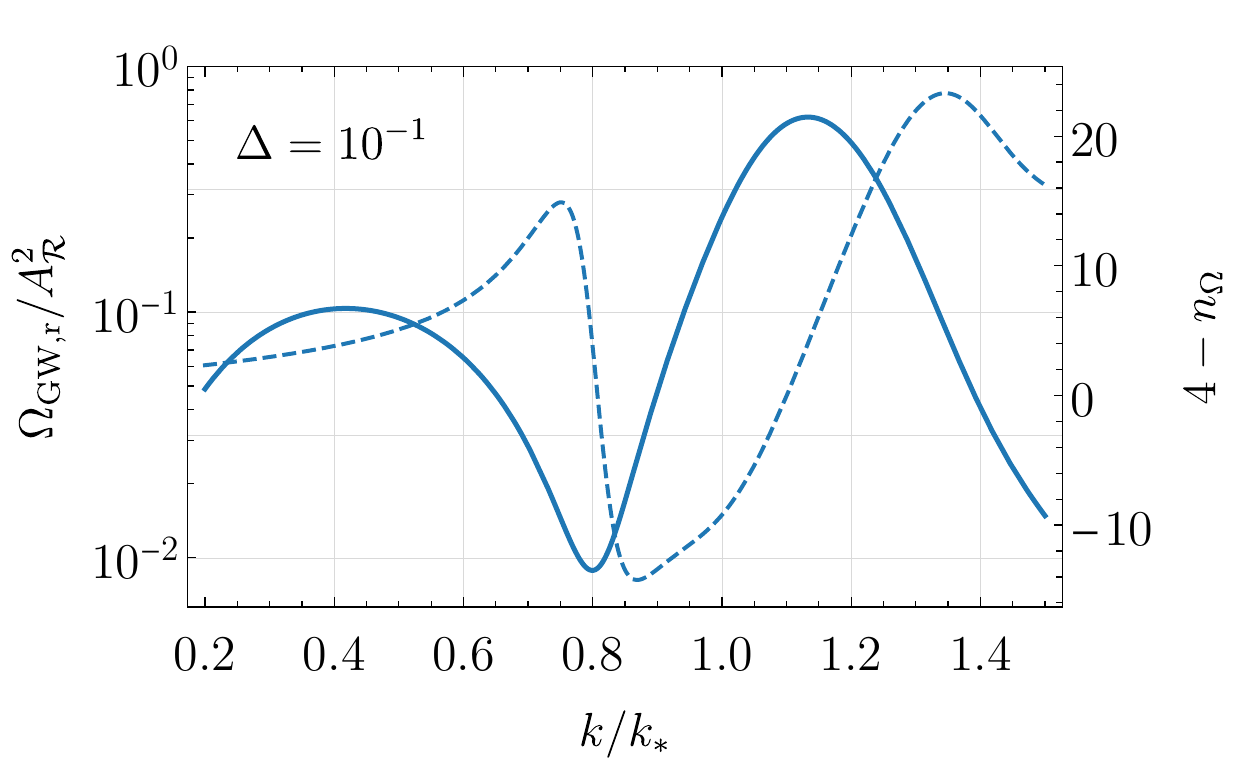}&\includegraphics[width=0.45\linewidth]{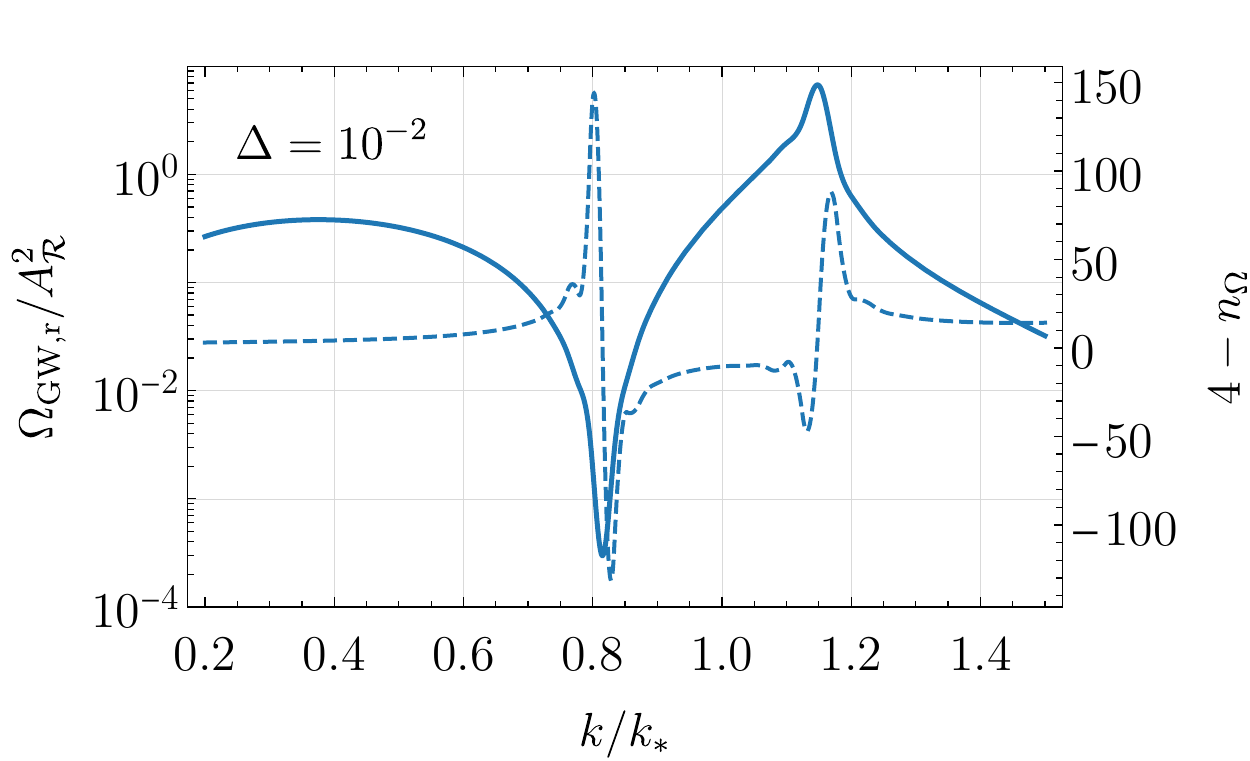}\\
    \includegraphics[width=0.45\linewidth]{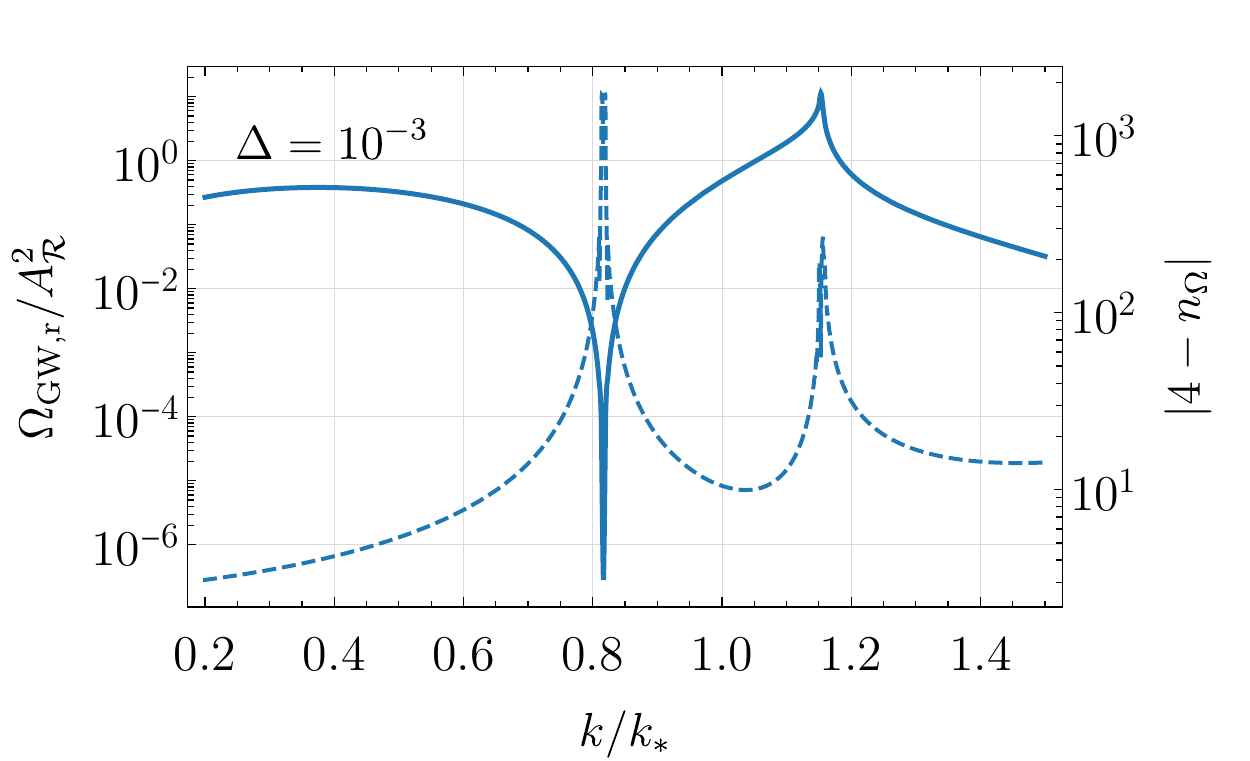}&\includegraphics[width=0.45\linewidth]{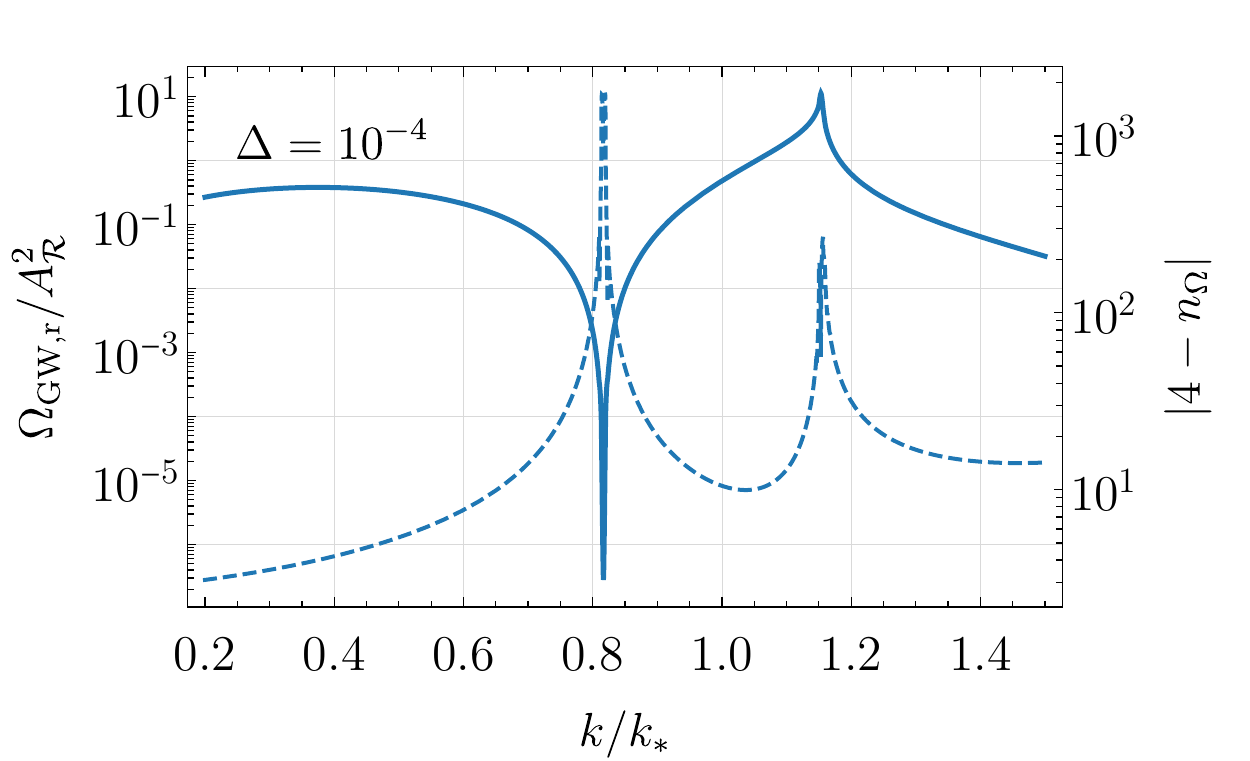}\\
\end{tabular}
     \caption{The dashed curves show the spectral tilt factor $4-n_{\Omega}$, the solid curves show the GW amplitude $\Omega_{\rm GW,r}/A_{\mathcal{R}}^2$.} 
    \label{fig:Omega_Delta}
\end{figure}
The GW spectrum and this spectral tilt factor $n_{\Omega}$ as a function of $k$ are plotted  in Fig.~\ref{fig:Omega_Delta} for different choices of $\Delta$. The frequency profile of the stochastic gravitational wave spectrum induced 
at second order from scalar fluctuations can be understood analytically,
at least for simple ans\"atze of the sourcing scalar spectrum. For example, for a delta-like scalar source, peaked at momentum $k \equiv k_*$, the resulting induced SGWB spectrum has a dip and goes to zero for $k/k_\star = \sqrt{2/3}$,
and a pronounced resonant peak at  $k/k_\star = \sqrt{4/3}$.

If the source scalar spectrum profile is narrow but not a delta-function, then the previous features are smoothed. Fig.~\ref{fig:Omega_Delta} shows the SGWB induced by a log-normal peak, for different choices of the characteristic width. The technical reason for these features can be found in the convolution integrals that give the SGWB at second order: see e.g. \cite{Ananda:2006af,Baumann:2007zm,Saito:2010pbh} for important early works on the subject, as well as \cite{Domenech:2021ztg} for a recent systematic review.~The features are  due to interference and resonant effects among the  scalar modes that source the GW background. 

The spectrum is normalised against $A_\rc^2$ because the latter quantity always results as overall   coefficient in front of the induced    $\Omegagw$ at second order (see also   \cite{Ananda:2006af,Baumann:2007zm,Saito:2010pbh,Domenech:2021ztg}). It is customary to plot the GW frequency profile singling out  such overall factor.

\subsubsection{Inflationary models with peaked spectra}
A number of inflationary models can produce a sufficiently narrow peak in the GW spectrum which can enhance the SGWB anisotropies by a factor of $\oc(100)$ or more.  Within single-field models, the curvature power spectrum has bounds in its growth rate (see e.g. \cite{Byrnes:2018txb,Carrilho:2019oqg,Ozsoy:2019lyy,Tasinato:2020vdk}) which restricts the steepness of the induced GW spectrum \cite{Matarrese:1997ay,Ananda:2006af,Baumann:2007zm}, and -- consequently --
 how large $n_{\Omega}$ can be. In fact, we find that single-field scenarios can in general be well-represented by the $\Delta=10^{-1}$ case of Fig.~\ref{fig:Omega_Delta}. On the other hand, models with modified gravity \cite{Pi:2017gih}, multiple fields \cite{Garcia-Bellido:1996mdl,Frampton:2010sw,Palma:2020ejf,Fumagalli:2020adf,Braglia:2020eai}, parametric resonance \cite{Chen:2019zza,Cai:2019jah,Cai:2019bmk,Chen:2020uhe} or particle production \cite{Fumagalli:2020nvq} may produce the narrow spectra corresponding to scenarios with $\Delta\leq 10^{-2}$. One should also mention models with non-standard kinetic terms, these too can support
an enhanced GW spectrum, see e.g. \cite{Mylova:2018yap,Ozsoy:2019slf}.
%  \replyref{``Aren’t there models with non-standard kinetic
% terms that are capable of inducing an enhancement? If yes, why aren’t
% they mentioned?"}{...}

\subsubsection{Other sources of peaked GW spectra}
\label{3.1.2}
GW from non-inflationary sources, e.g. phase transitions \cite{Caprini:2015zlo}, cosmic strings  \cite{Auclair:2019wcv}, can also produce peaked spectra although the enhancement from the tilt is typically of the same order as for a power-law spectrum. As an example, let us consider the SGWB produced from cosmological phase transitions. For these backgrounds, the sound wave contribution is generally larger than the other contributions and has a steeper spectral shape given by \cite{Mazumdar:2018dfl}
\begin{align}
    \label{eq:PT_shape}
    S(k,k_*)=\left(\frac{k}{k_*}\right)^3\left(\frac{7}{4+3(k/k_*)^2}\right)^{7/2}\,,
\end{align}
with 
\begin{align}
    \label{eq:PT_Omegagw}
    \Omegagw(k) = \Omegagw(k_*)S(k,k_*)\,,
\end{align}
where $\Omegagw(k_*)$ is the GW amplitude at the peak wavenumber $k_*$. From Eq.~\eqref{eq:PT_shape} one can infer that the factor $(4-n_{\Omega}(k))$ appearing in the R.H.S. of Eq.~\eqref{eq:def_deltagw} is \oc(1). Thus, the enhancement in this case is much smaller than what GW sourced from peaks in the curvature power spectrum allow for. 
In the next section, we focus on the latter scenario and explore the observational consequences through an example of SIGW detectable at mHz scales.

\subsection{LISA GW-PBH scenario}
\label{sec:LISA_GW}
\begin{figure}
\begin{minipage}{0.49\textwidth}
    \centering
    \includegraphics[width=0.95\linewidth]{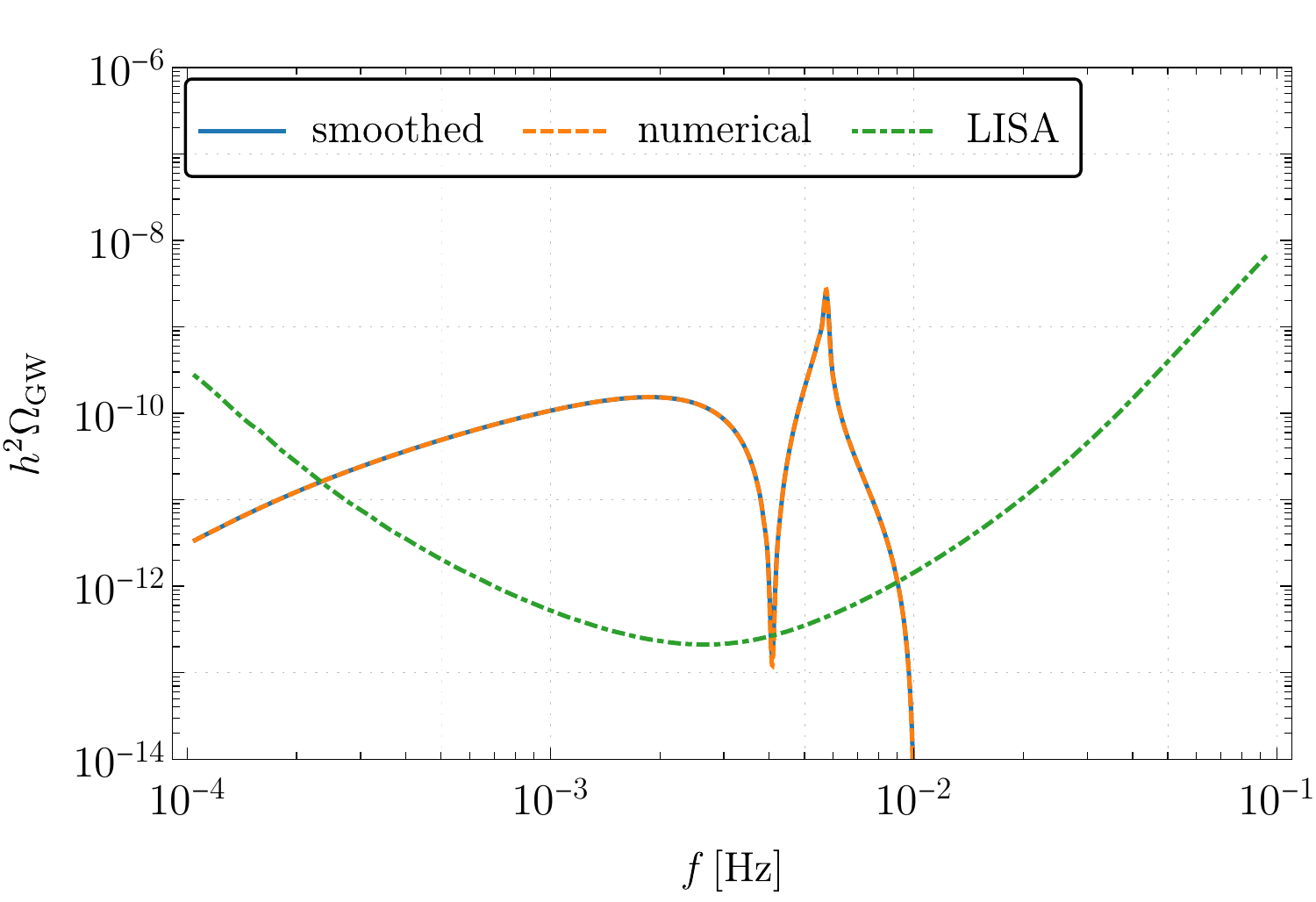}
\end{minipage}
\begin{minipage}{0.49\textwidth}
    \centering
    \includegraphics[width=0.95\linewidth]{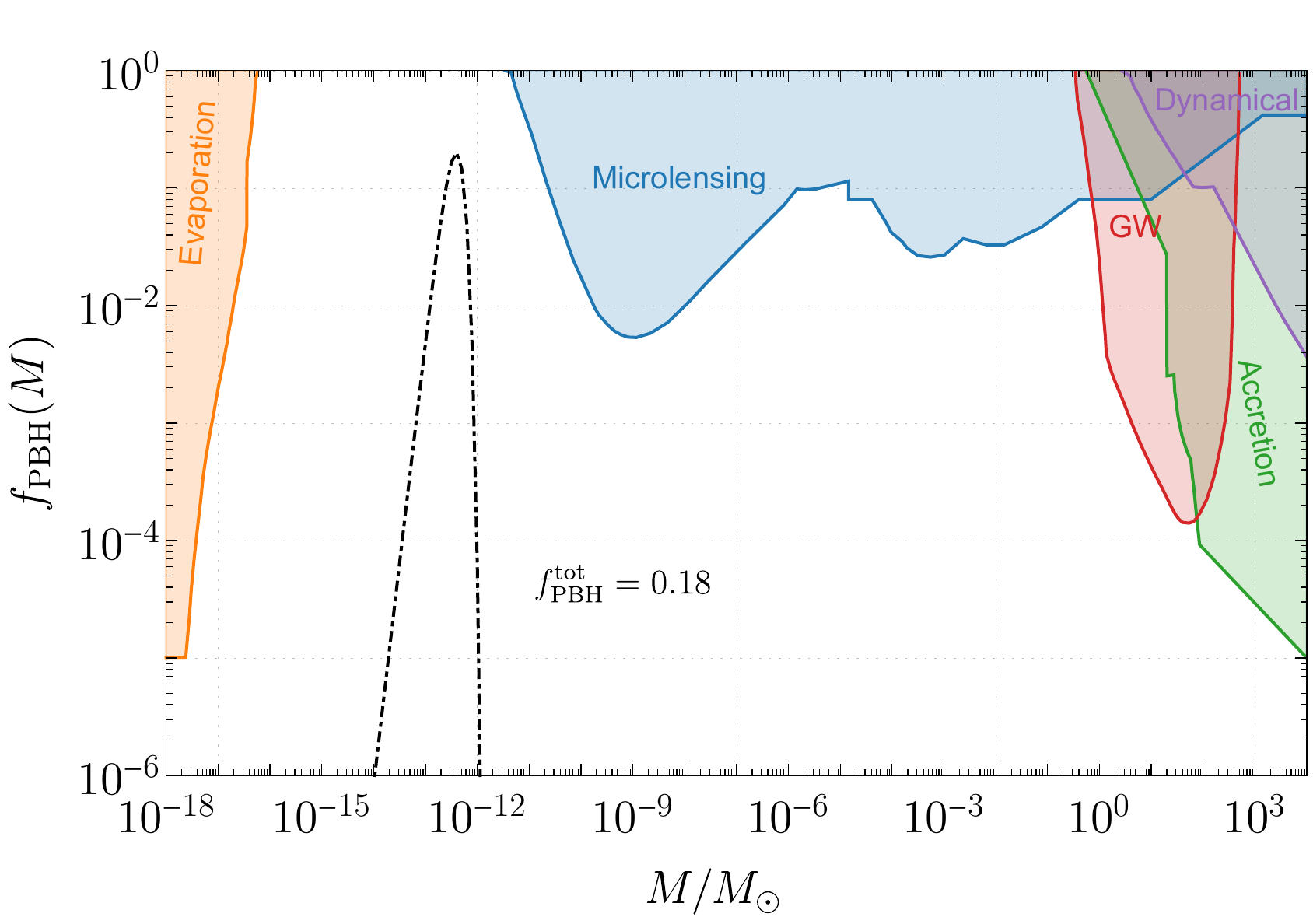}
\end{minipage}
    \caption{Left - The induced GW spectrum at LISA scales for the parameter choice given in table~\ref{tab:param_LISA} and the smoothed spectrum defined in Eq.~\eqref{eq:smoothed_GW}. The green dashed curve shows power-law integrated sensitivity curve for LISA \cite{Smith:2019wny} (see \cite{Lewicki:2021kmu} for the updated sensitivity curves due to the impact of foregrounds and resolvable binaries). Right - The PBH abundance for the same values of the parameters plotted along with the constraints from microlensing \cite{Smyth:2019whb,EROS-2:2006ryy,Croon:2020ouk,Griest:2013aaa,Oguri:2017ock,Niikura:2019kqi,Macho:2000nvd}, accretion \cite{Manshanden:2018tze,Lu:2020bmd,Serpico:2020ehh,Hektor:2018qqw}, GW \cite{Hutsi:2020sol}, evaporation \cite{Carr:2009jm,Clark:2016nst,Boudaud:2018hqb,Clark:2018ghm,Dasgupta:2019cae,DeRocco:2019fjq,Laha:2019ssq,Laha:2020ivk,Laha:2020vhg,Saha:2021pqf,Mittal:2021egv} and dynamical constraints \cite{Monroy:2014,Brandt:2016aco}.}
    \label{fig:lisa_gw_pbh}
\end{figure}

As a representative example, we consider a scenario in which the scalar induced GW spectrum can be detected by LISA \cite{amaroseoane2017laser}. For this, we take the peak of the log-normal power spectrum of Eq.~\eqref{eq:lgnormal_pk} to correspond to a frequency within the LISA band $10^{-4}\text{ Hz}\leq f \leq 10^{-1} \text{ Hz}$. The relation between frequency and wavenumber is given by
\begin{align}
    \frac{k}{\text{Mpc}^{-1}} \simeq 6.5\times 10^{14} \frac{f}{\text{Hz}}\,.
\end{align}
The induced GW spectrum is plotted in the left panel of Fig.~\ref{fig:lisa_gw_pbh} for the parameter choice of table~\ref{tab:param_LISA}. Note that to avoid numerical artifacts arising in the calculations we have also smoothed the $\Omegagw$ spectrum near the peak using the peak width as the smoothing scale. The smoothed spectrum $\tlgw$ is defined as
\begin{align}
    \label{eq:smoothed_GW}
    \tlgw(f) = \frac{1}{2\Delta}\int_{f e^{-\Delta}}^{f e^{\Delta}}\Omegagw(f')\,d\ln f'\,.
\end{align}
% \replyref{ ``Why is it necessary to introduce the smoothed
% spectrum? Are numerical artifacts unavoidable in this computation?" }{ Appears to be the case, the integrand has highly oscillatory terms. I also checked with the new code\footnote{\url{https://github.com/Lukas-T-W/SIGWfast}} from Guillem's collaborator Lukas, even with that it is not easy to avoid numerical issues especially for $\Delta\ll 1$. Secondly, frequency resolution of interferometers, which for anisotropies will be worse than the frequency resolution obtained simply from $\delta f \sim 1/T_\mathrm{obs}$. -AM}

{\renewcommand{\arraystretch}{1.25}%
\begin{table}
    \centering
    \begin{tabular}{|c|c|}
    \hline
    $\Delta$ & $10^{-2}$\\
    \hline
     $f_*$ & $5\times 10^{-3}\, \rm Hz$    \\
     \hline
     $k_*$ & $3\times 10^{12}\, \rm Mpc^{-1}$\\
     \hline
     $A_{\rc}$ & $7.5\times 10^{-3}$\\
     \hline
\end{tabular}
    \caption{Parameter choice for the LISA GW scenario.}
    \label{tab:param_LISA}
\end{table}}
The enhancement of the curvature power spectrum can also lead to the production of PBH and one ought to ensure that the scenario considered here is well within the region allowed by current constraints on the PBH abundance. The relation between PBH mass at formation and the comoving scale $k$ re-entering the horizon (during the radiation era) can be written as \cite{Garcia-Bellido:2017aan,Sasaki:2018dmp},
\begin{align}
   \mpbh \simeq 30\msun\left(\frac{\gamma}{0.2}\right)\left(\frac{g_{*}}{10.75}\right)^{-1/6}\left(\frac{2.9\times 10^5 {\rm Mpc^{-1}}}{k}\right)^2\,.
\end{align}
For peak frequencies/wavenumbers relevant for LISA this leads to PBH with masses in the range $\mpbh \sim \mathcal{O}(10^{-15}\text{ -- }10^{-12})\msun$ (see \cite{Bartolo:2018evs,Bartolo:2018rku} for a detailed analysis of the PBH-GW scenario for LISA). In our case the corresponding PBH abundance is shown in the right panel of  Fig.~\ref{fig:lisa_gw_pbh} for $\Delta=10^{-2}$ along with the corresponding constraints, compiled using the tool provided in \cite{bradley_j_kavanagh_2019_3538999}. To calculate the mass function of PBH we have followed the method of ref. \cite{Byrnes:2018clq} which uses a simple Press-Schechter approach and accounts for the effects of critical collapse\footnote{A detailed comparison of different methodologies for calculating the PBH abundances and mass distribution can be found in \cite{Gow:2020bzo}.}. The PBH mass function $\fpbh(M)$ is defined as,
\begin{align}
    \frac{\Omega_{\rm PBH}}{\Omega_{\rm DM}} = \int \fpbh(M)\, d\ln M\,,
\end{align}
where the L.H.S. denotes the total fraction of the dark matter density constituted by PBH. Note that the constraints shown in Fig.~\ref{fig:lisa_gw_pbh} are strictly valid only for a monochromatic mass function and for extended mass functions the constraints need to be calculated differently \cite{Carr:2017jsz}. 

\subsubsection{Angular power spectrum of GW anisotropies}
\label{sec:angular_spectra}
\begin{figure}
\centering
\begin{tabular}{c c}
     \includegraphics[width=0.45\linewidth]{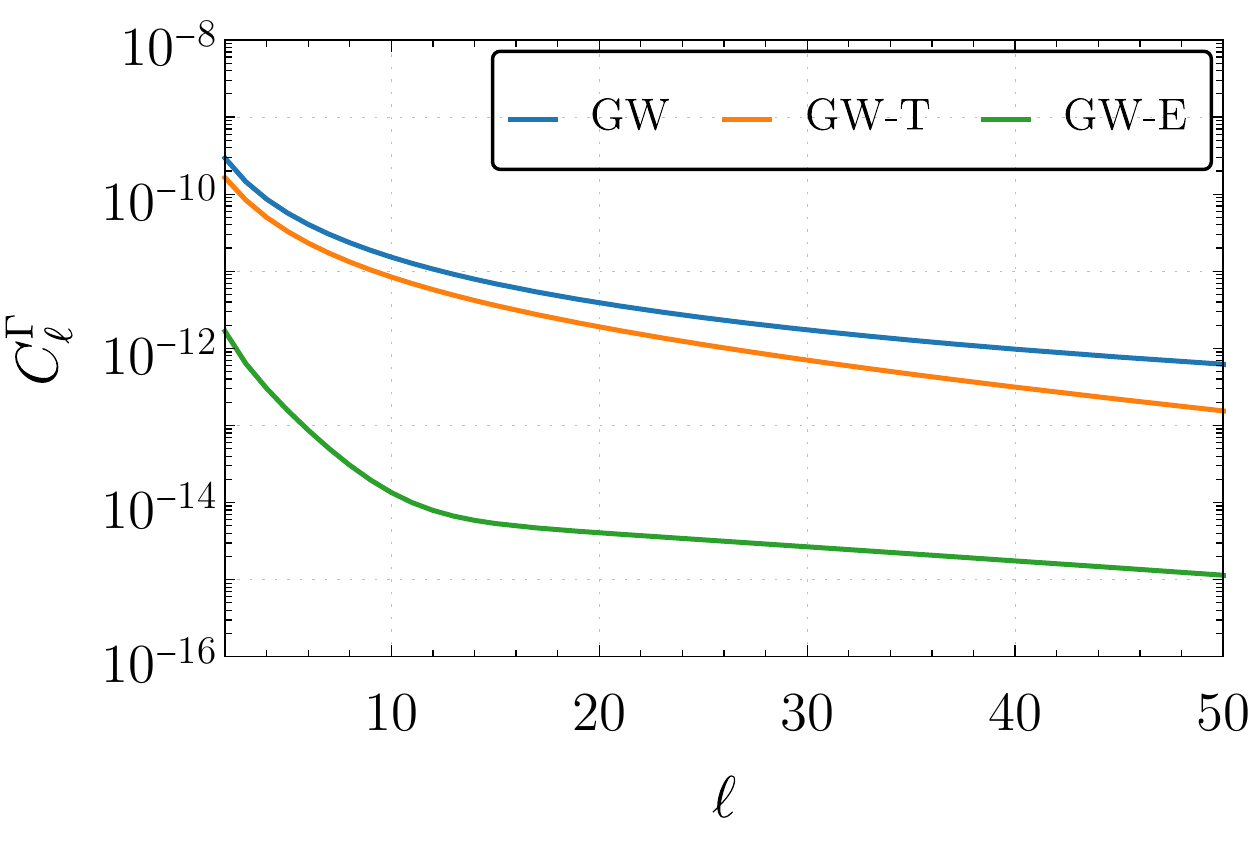}&
     \includegraphics[width=0.45\linewidth]{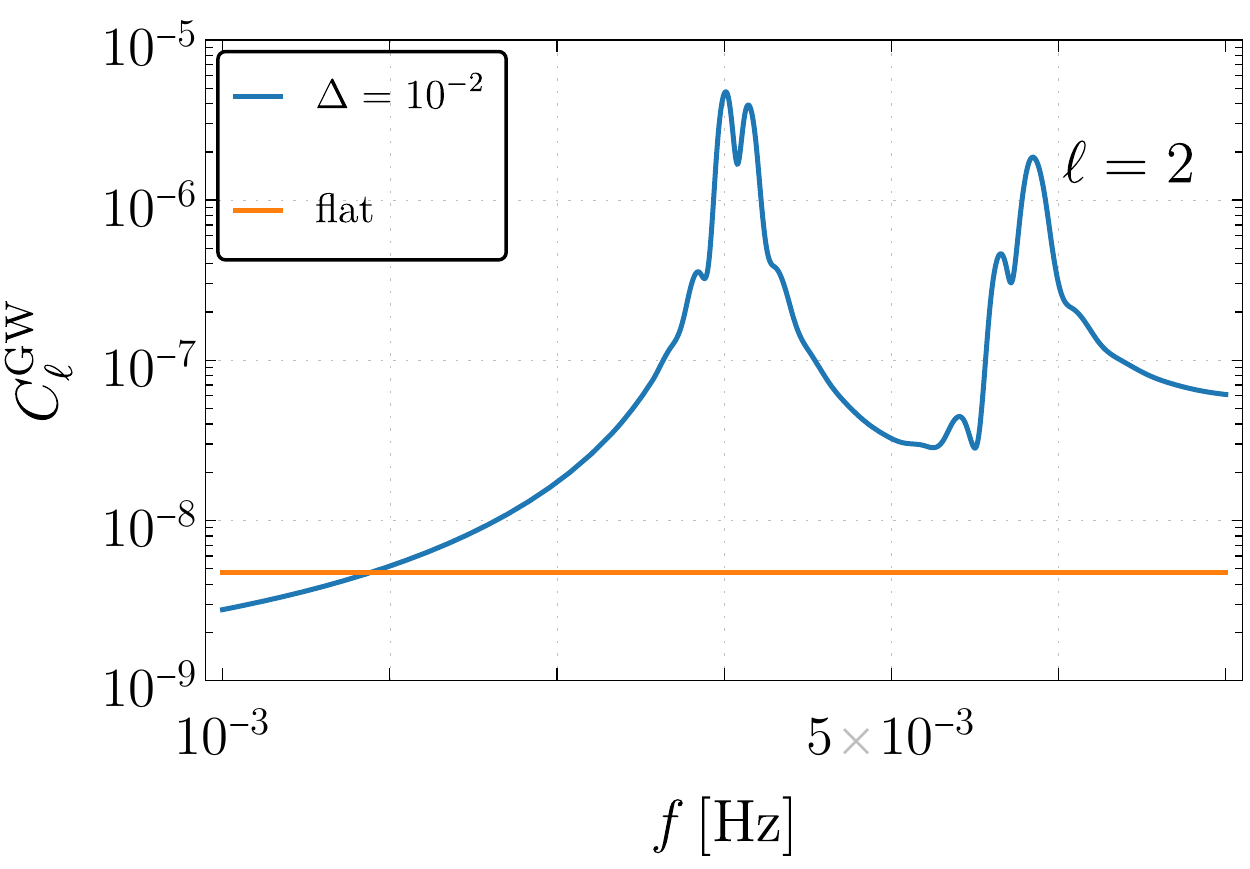}\\
\end{tabular}
    \caption{Left - Angular power spectrum of the GW anisotropies and their cross-correlation with CMB-T, E. Right - Frequency dependence of the angular power spectrum, for $\ell=2$, for the LISA model of table~\ref{tab:param_LISA} (blue line) and for a flat spectrum (orange line) plotted in a frequency range where LISA has the highest sensitivity.}
    \label{fig:clgw_freq}
\end{figure}

We now calculate the angular power spectrum of the GW anisotropies for the LISA GW scenario of table~\ref{tab:param_LISA}. The angular power spectrum is defined as 
\begin{align}
\label{eq311}
    \langle \Gamma_{ \ell m} \Gamma_{ \ell' m'} \rangle =  \delta_{\ell \ell'}\delta_{m m'}C_{\ell}^{\Gamma}\,,
\end{align}
where we have assumed statistical isotropy. We denote by $C_{\ell}^{\rm GW}$ the angular power spectrum of $\deltagw$ which is related to the above quantity as follows,
\begin{align}
\label{eq312}
    C_{\ell}^{\rm GW}(k) = \left(4 - n_{\Omega}(k)\right)^2  C_{\ell}^{\Gamma}\,,
\end{align}
with $n_{\Omega}$ given by Eq.~\eqref{eq:def_deltagw}. In the left panel of Fig.~\ref{fig:clgw_freq} we plot the autocorrelation of the GW anisotropies and their cross-correlation with CMB temperature and E-mode polarisation anisotropies. We obtained the T and E transfer functions from CAMB \cite{Lewis:1999bs}. Note that the propagation anisotropies of the CGWB quantity $\Gamma$ (Eq.~\eqref{eq:GW_ad}), and consequently $C_\ell^\Gamma$, are frequency independent.
In the right panel, we show the frequency dependence of $C_{\ell}^{\GW}$ (for $\ell=2$) for the LISA GW scenario and for a flat  spectrum of GW for comparison. For simplicity, we have only considered the SW contribution for GW anisotropies, Eq.~\eqref{eq:GW_ad}, since this is the dominant contribution on large angular scales, and ignored the ISW term. These scales are the most relevant for the GW detectors under consideration here since these are limited by their angular resolution $\ell_{\rm max}\sim 15\text{--}30$  \cite{Kudoh:2004he,Alonso:2020rar,Contaldi:2020rht}.  It is worth noting that this frequency dependence of the anisotropies arises for any SGWB that is different from a power-law ($n_\Omega=\rm const.$), whereas $C_\ell^\Gamma$ for the anisotropies from propagation is expected to be the same for all CGWB.

\subsubsection{SNR of the anisotropies}
\begin{figure}
\centering
\begin{tabular}{c c}
     \includegraphics[width=0.45\linewidth]{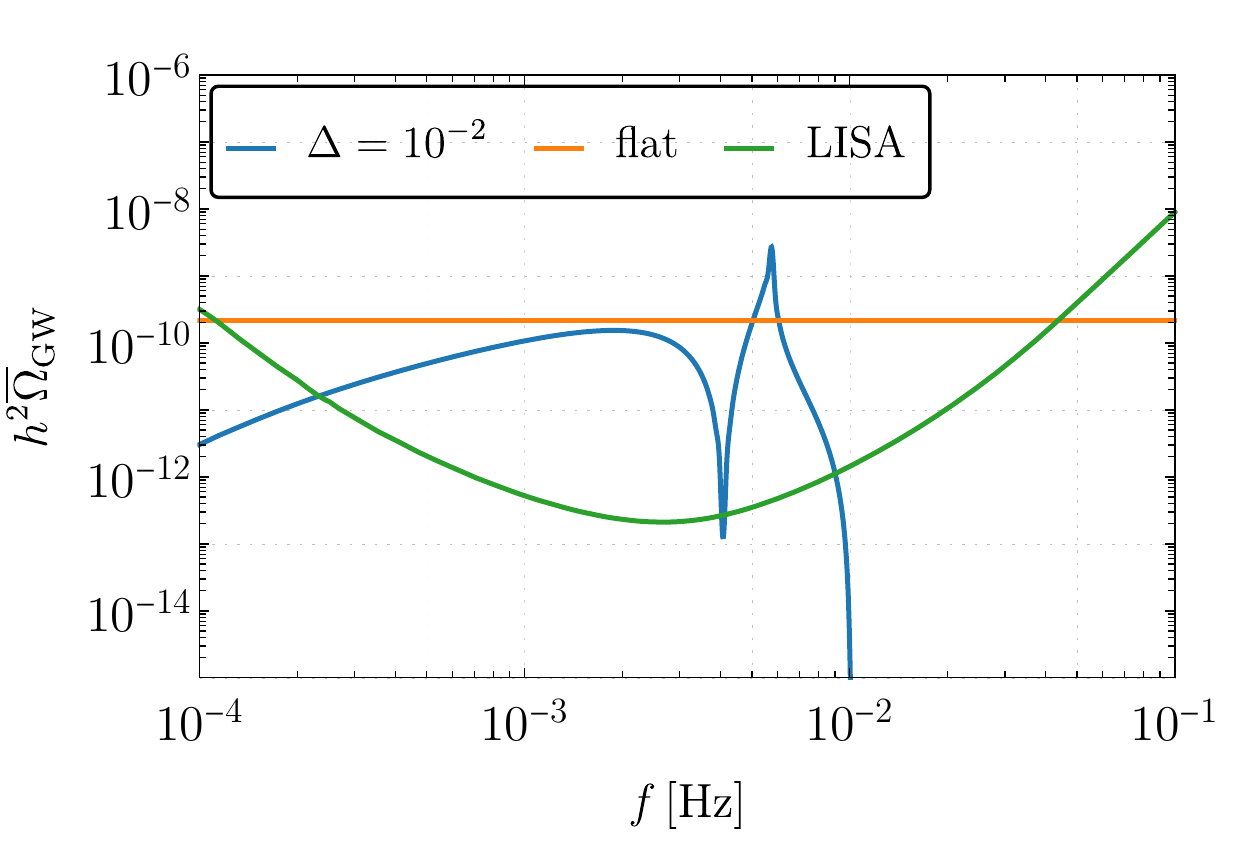}& \includegraphics[width=0.45\linewidth]{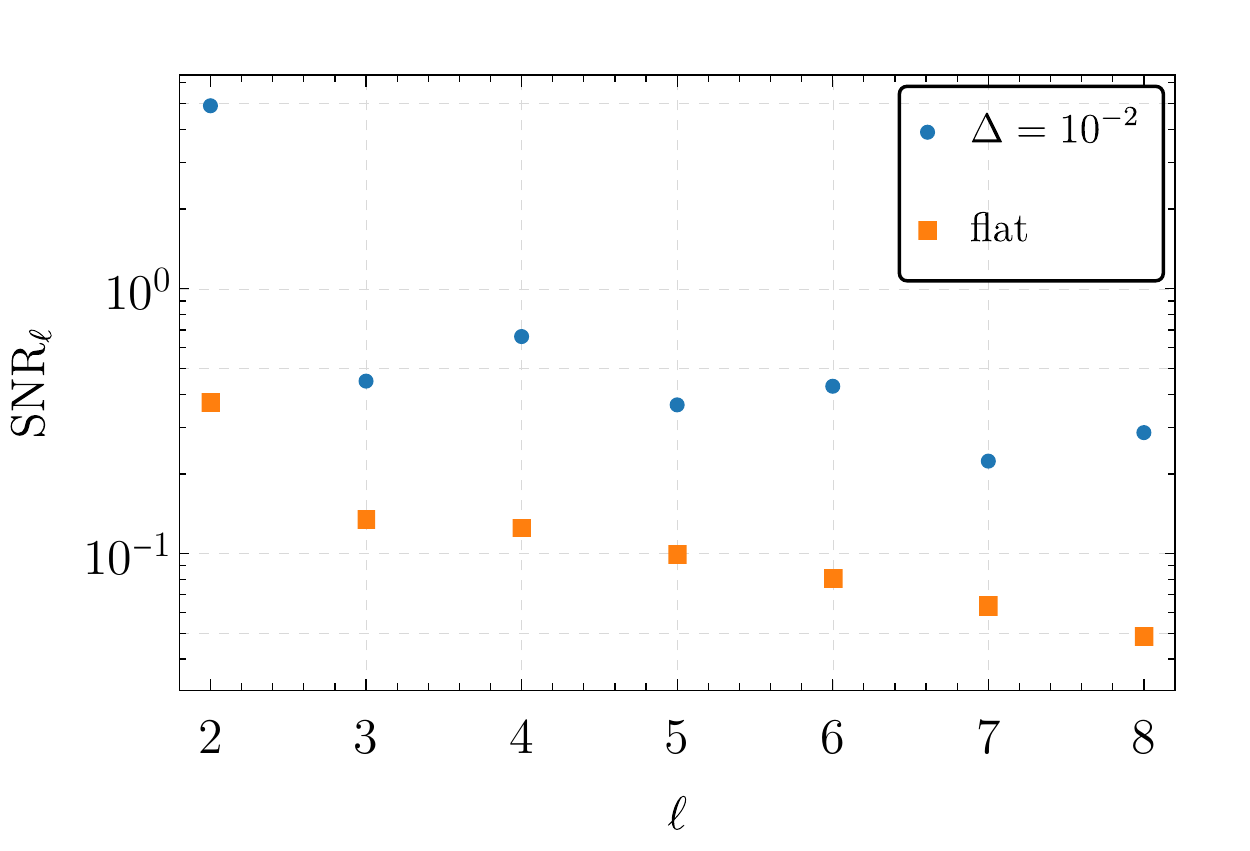}\\
\end{tabular}
    \caption{Left - $\Omega_\GW(f)$ plotted for the two spectra. Right - SNR of the individual multipoles with $T_{\rm obs}=3\rm\,years$.}
    \label{fig:equiv_snr_comp}
\end{figure}

In this section we show that the enhancement of anisotropies for a peaked spectrum can make these anisotropies easier to detect compared to the case of a standard power-law spectrum. A related phenomenon was recently pointed out in \cite{Cusin:2022cbb} for the case of kinematic anisotropies.
For the purpose of this comparison, we consider a flat power-law spectrum with an amplitude such that the resulting SNR of the monopole is the same in the LISA range. We have verified that the results for a power law with a small non-zero spectral tilt $n_\Omega\sim \oc(1)$  are similar so we do not show them separately. 
We plot the SNR for the individual multipoles in Fig.~~\ref{fig:equiv_snr_comp}. Our results indicate that detecting the anisotropy is easier for a peaked spectrum than for a flat spectrum. 
We see that, although the $\ell=2$ is detectable with a LISA-Taiji network for the peaked spectra, it is not detectable for the flat spectra. 
The multipole SNR is defined as \cite{Bartolo:2022pez}
\begin{align}
\label{eq:snr_def}
    \SNR_{\ell}^2 = \int df\, C_{\ell}^{\GW}(f)\left(\frac{\Omegagw(f)}{\Omega^{\ell}_{\GW,n}(f)}\right)^2\,,
\end{align}
where the quantity $\Omega^{\ell}_{\GW,n}(f)$ is the effective angular sensitivity of the detector network to the $\ell-$th multipole. The quantity $\Omega_{\rm GW,n}^{\ell}(f)$ is defined as 
\begin{align}
    \Omega_{\rm GW,n}^{\ell}(f)^{-1} \equiv T_{\rm obs}\sum_{AB}\left(\frac{2}{5}\right)^2\left(\frac{4\pi^2 f^3}{3H_0^2}\right)^{-2} \frac{1}{N^A_f N^B_f}\frac{\sum_{m}|\mathcal{A}_{AB}^{\ell m}(f)|^2}{(2\ell+1)},
\end{align}
where $T_{\rm obs}$ is the time of observation, $N_f^A$ denotes the noise Power Spectral Density for detector $A$ and $\mathcal{A}^{\ell m}_{AB}$ the spherical harmonic transform of the antenna pattern for the detector pair $AB$.

\begin{figure}
    \centering
      \includegraphics[width=0.45\textwidth]{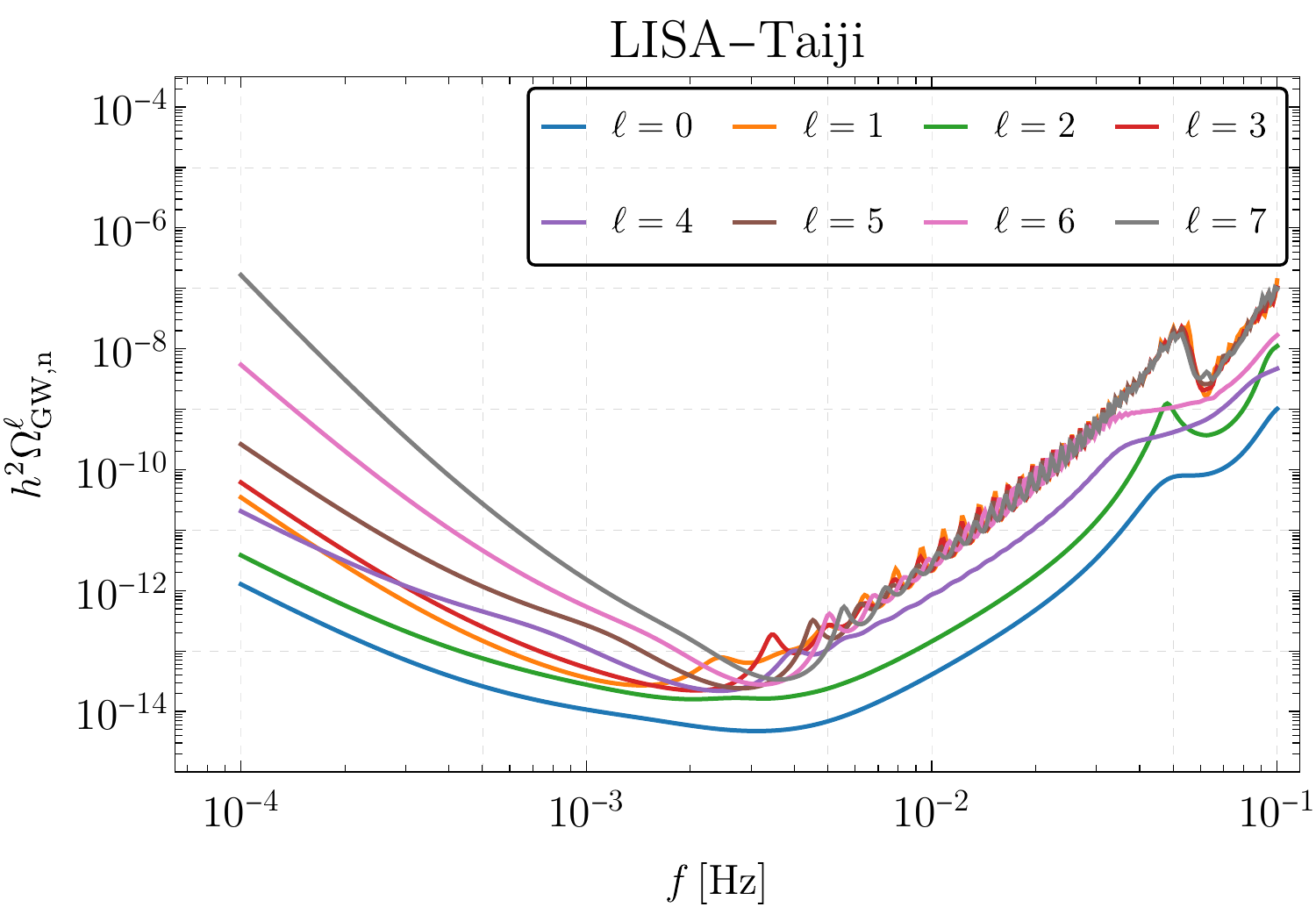}
      \includegraphics[width=0.45\textwidth]{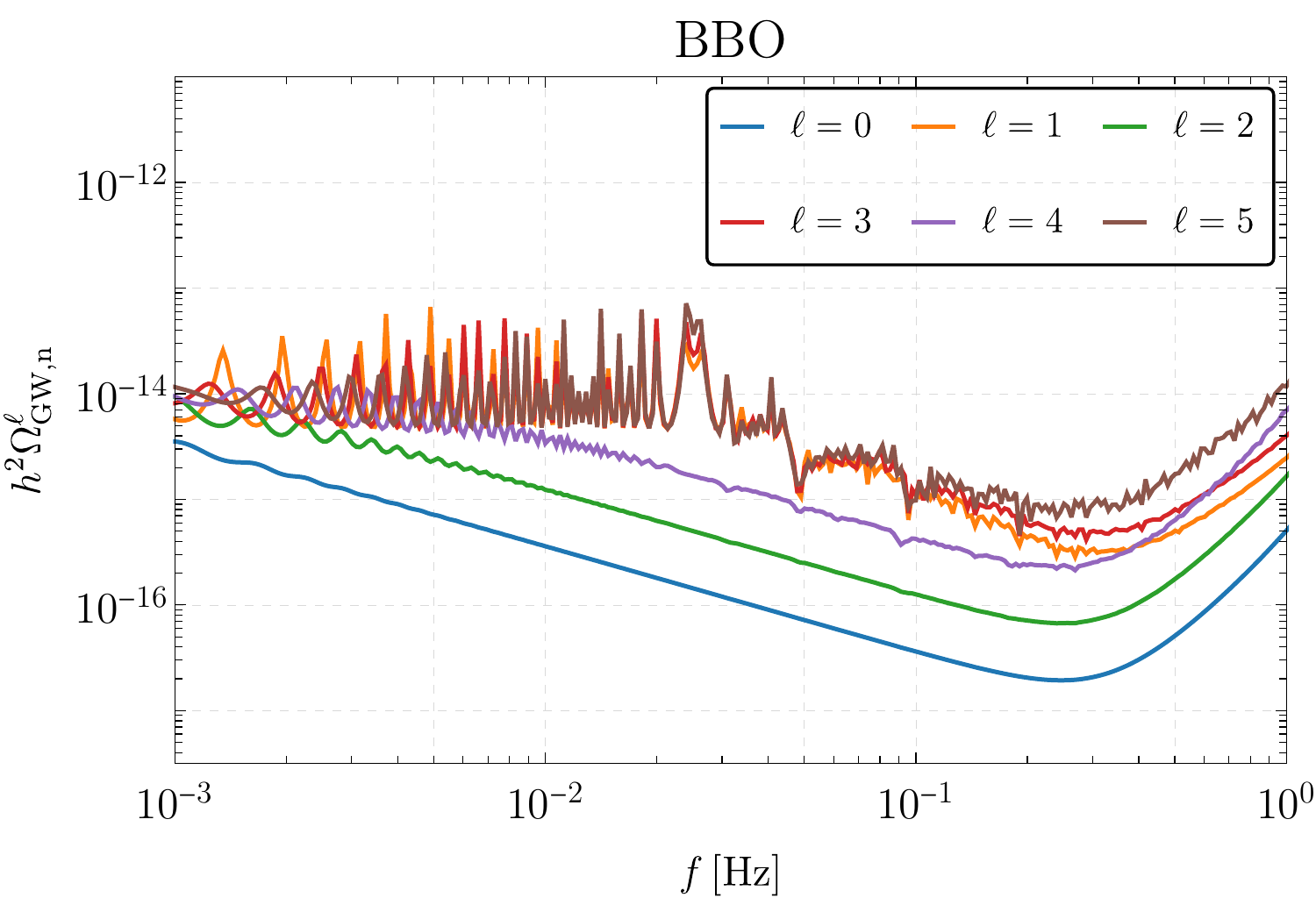}
    \caption{Sensitivity of LISA-Taiji and BBO to the different multipoles of the SGWB as a function of frequency. The time of observation is taken to be 3 years for both plots.} 
    \label{fig:nl_lisa}
\end{figure}

We use the code schNell\footnote{\url{https://github.com/damonge/schNell} (see also the companion work in \cite{Alonso:2020rar}).} to compute $\Omega_{\rm GW,n}^{\ell}(f)$.
Fig.~\ref{fig:nl_lisa} contains a plot of the quantity $\Omega_{\rm GW,n}^{\ell}(f)$ for LISA-Taiji \cite{Ruan:2020smc} and for the BBO configuration with 4 constellations \cite{Crowder:2005nr}. We include all LISA-Taiji cross-correlations as well correlations internal to both LISA and Taiji while for BBO we only consider cross-correlations among the different vertices. The addition of multiple detectors separated by large distances improves angular sensitivity compared to a single LISA-like constellation, e.g. see Fig.~9 of ref.~\cite{Bartolo:2022pez} or Fig.~6 of ref.~\cite{Kudoh:2004he}. The noise curves for the various detectors can be found in \cite{Smith:2019wny} for LISA, \cite{Ruan:2020smc} for Taiji and \cite{Smith:2016jqs} for BBO.

\subsubsection*{SNR of the cross-correlation}
\begin{figure}
    \centering
    \includegraphics[width=0.45\linewidth]{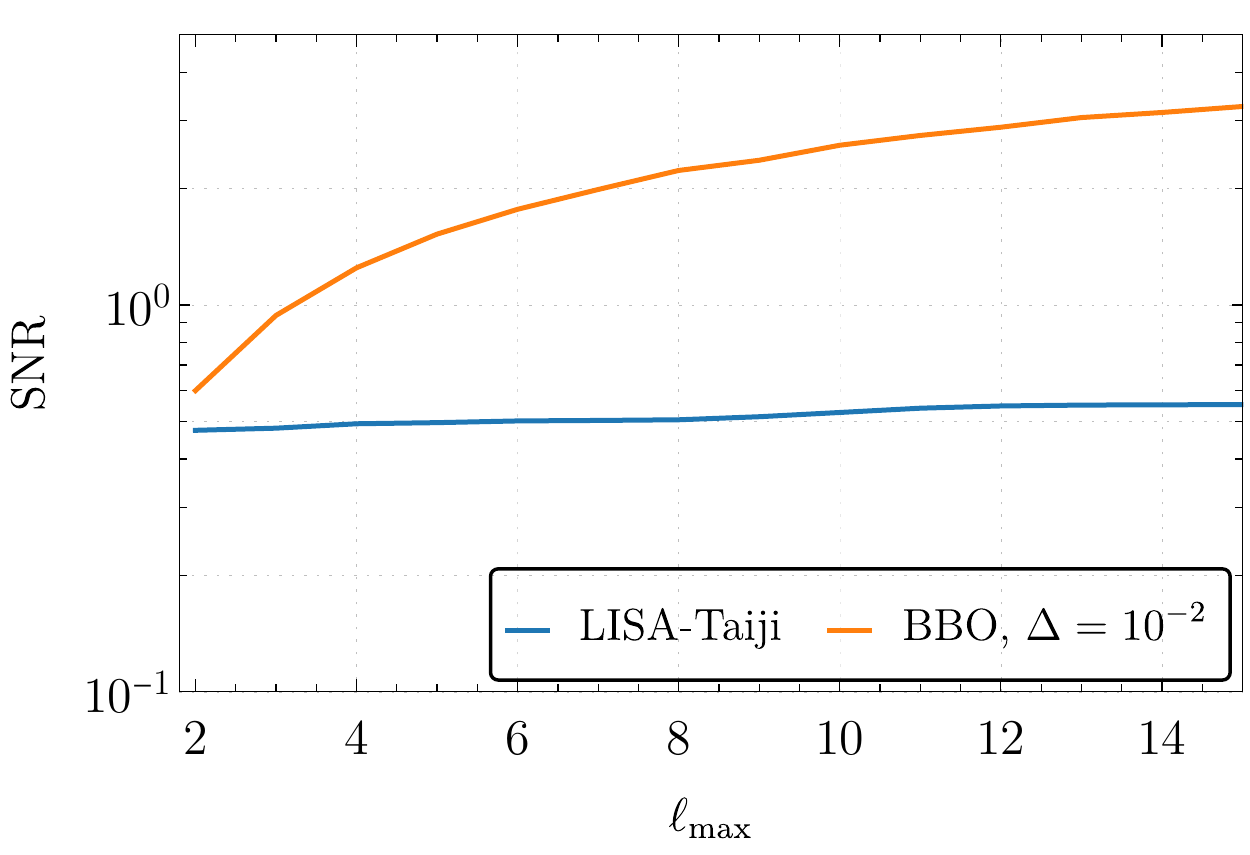}
    \caption{SNR of the cross-correlation between the GW anisotropies and the CMB-T, E mode anisotropies.}
    \label{fig:snr_cross}
\end{figure}
The SNR of the cross-correlation with the CMB-T, E anisotropies is defined as\footnote{Note that incomplete sky coverage can lead to correlations between nearby $\ell$ thus degrading the sensitivity to the individual multipoles. As in ref.~\cite{Ricciardone:2021kel}, we have assumed for simplicity that the GW and CMB maps used to calculate the cross-correlation are full sky which makes our estimate an optimistic one. A reduction of the SNR by a factor $\sqrt{f_\mathrm{sky}}<1$ would be expected in a more realistic case~\cite{Tegmark:1997vs,Tegmark:1999ke}.}
\begin{align}
  \text{SNR}^2 = \sum_{\ell=2}^{\ell_{\rm max}}\sum_{\rm X=T,E}(2\ell+1)\frac{\left(C^{\rm{X}\Gamma}\right)^2}{\left(C^{\rm{X}\Gamma}\right)^2+\left(C_{\ell}^{\Gamma}+N_{\ell}^{\Gamma}\right)C_{\ell}^{\rm X}}\,,
\end{align}
where the quantity $N_\ell^\Gamma$ is (for $\ell>0$)
\begin{align}
    \label{eq:Nell_gamma}
    N_{\ell}^{\Gamma} \equiv \left[\int df \frac{(4-n_{\Omega}(f))^2\Omegabar(f)^2}{\Omega_{\rm GW,n}^{\ell}(f)^2}\right]^{-1}\,.
\end{align}
We plot the SNR in Fig.~\ref{fig:snr_cross} as a function of $\ell_{\rm max}$ for observation by LISA-Taiji and BBO. For the representative example considered here, this cross-correlation will be detectable by a BBO-level GW detector but not with LISA-Taiji. The GW-T cross-correlation provides the entirety of the cross-correlation SNR while the GW-E contribution is negligible.

\section{Conclusion }
\label{sec:conclusion}

These are very exciting times for GW astronomy: observations of compact object mergers by LIGO-Virgo and KAGRA have already demonstrated its tremendous potential to shed light on stellar and black hole physics, late-time cosmology as well as general relativity itself. \\
\indent The detection of a background resulting from the superposition of a large number of these astrophysical sources is also expected to happen in the near future. When it comes to a cosmological background, such as the one from inflation, detection will likely be more challenging and require space-based and future ground-based interferometers. The importance of detecting cosmological backgrounds cannot be understated. These are generated well before the CMB and thus provide access to the primordial universe at much earlier times than what is possible by electromagnetic observations alone.\\

\indent Much like the CMB, the CGWB is predicted to be nearly isotropic with small fluctuations that may be inherent to the production mechanism itself or may arise due to propagation in a universe that is not perfectly homogeneous and isotropic.\\

\indent In the first part of this paper we have reviewed two different approaches to derive these ``propagation'' anisotropies and shown how, under some simplifying assumptions, they can be related to each other once a careful choice of the GW initial conditions is performed. \\
\indent We also pointed out how the adiabatic or isocurvature nature of the primordial perturbations affects this anisotropy through the intrinsic density fluctuation at the time when the GW start propagating. Whether or not a significant amplitude of isocurvature perturbations could be produced to noticeably alter the SGWB anisotropies while leaving the CMB unaffected is a subject that merits further investigation. We plan to address such matters in an upcoming work. \\

In the second part of this paper, we analysed the propagation anisotropy spectrum of SGWB sourced at second order in the curvature perturbation. We saw that if the primordial curvature power spectrum is sharply peaked around some scales, as is possible in scenarios of PBH production, the SGWB anisotropies can be enhanced by a factor $\sim \mathcal{O}(10 \text{--}100)$ relative to those of a power-law SGWB spectrum. This enhancement arises due to the fact that the multiplicative factor $(4-n_{\Omega})$ in the definition of $\deltagw$ in Eq.~\eqref{eq:def_deltagw} becomes large if the SGWB spectrum is sharply peaked. \\
\indent Through the representative example of a scalar induced SGWB detectable at mHz frequencies, we showed that even though this enhancement only affects certain scales, it can nevertheless make these anisotropies easier to detect relative to those of a standard power-law spectrum. For this representative example, we demonstrated the capability of a LISA-Taiji network to detect the $\ell=2$ multipole as well as that of BBO to detect the cross-correlation with CMB anisotropies. Larger values of the SGWB monopole would naturally increase the detectability of these anisotropies, as would the improved sensitivity and angular resolution of futuristic detector networks like BBO/DECIGO.\\ \indent We also considered SGWB from other sources which may exhibit a peaked spectral shape. We find that in a representative case (see section \ref{3.1.2}) the enhancement of the anisotropies is small compared to the PBH scenario mentioned above. Irrespective of whether the enhancement is significant or not, the distinct frequency dependence of the SGWB anisotropies for such spectra could possibly be exploited to separate them from the anisotropies associated with other CGWB or even SGWB of astrophysical origin. This idea has been previously used for the CMB to subtract the contribution from foreground sources. A dedicated analysis of its feasibility for GW is also worth pursuing and would bring us one step closer to the detection of these GW anisotropies.

\subsection*{Acknowledgments}

M.F. would like to acknowledge support from the “Atracci\'{o}n de Talento” grant 2019-T1/TIC15784, his work is partially supported by the Spanish Research Agency (Agencia Estatal de Investigaci\'{o}n) through the Grant IFT Centro de Excelencia Severo Ochoa No CEX2020-001007-S, funded by MCIN/AEI/10.13039/501100011033. GT is   partially  funded by the STFC grant ST/T000813/1. For the purpose of open access, the authors have applied a Creative Commons Attribution (CC BY) licence to any Author Accepted Manuscript version arising.

\appendix
\section{Boltzmann equation in \texorpdfstring{$\zeta$}{zeta}-gauge}
\label{app:Boltzmann_zeta}
In this Appendix, we provide additional details relevant to the calculations of Sec.~\ref{sec:Boltzmann_zeta} and then briefly review the original computation of ref.~\cite{Alba:2015cms}. 

To begin with, let $\lambda$ be the affine parameter along the graviton geodesic. The graviton 4-momentum is then $P^\mu=dx/d\lambda$. Let us express this in terms of the quantity $p$ defined as $p^2 \equiv g_{ij}P^i P^j $, and $\hn$, the unit vector along the GW direction. To this end, we first write $P^i = C\hn$, where $C$ can be determined in terms of $p$ as,
\begin{align}
    \label{eq:P^i_zeta}
    p^2 &= g_{ij} P^i P^j = a^2(1+2\zeta)C^2\delta_{ij}n^i n^j   \nonumber \\
   & \implies C = \frac{p}{a}(1-\zeta)\,.
\end{align}
Next, using $g_{\mu\nu}P^\mu P^\nu=0$ we find,
\begin{align}
    \label{eq:P^0_zeta}
    P^0 = \frac{p}{a}\left(1 - \frac{2}{5aH}\partial_i\zeta n^i\right)\,.
\end{align}
To solve the Boltzmann equation \eqref{eq:Boltzmann_eq} we need to calculate $dq/d\eta$ with $q=|\vec p|a$. Using Eqs.~\eqref{eq:P^i_zeta} and \eqref{eq:P^0_zeta} one finds $P_0 = -q$. Thus, it only remains to solve the geodesic equation for $P_0$. This is given by \cite{Mukhanov:2005sc},
\begin{align}
    \frac{dP_0}{d\lambda} = \frac{1}{2}\partial_0 g_{\alpha\beta}P^{\alpha}P^{\beta}\,.
\end{align}
Using the fact that $d/d\lambda = P^0 d/d\eta$ one can re-write the above equation as 
\begin{align}
    P^0\frac{dP_0}{d\eta} = \frac{1}{2}\partial_0 g_{\alpha\beta}P^{\alpha}P^{\beta}.
\end{align}
To simplify things further, we can work with the re-scaled metric without the scale factor which we denote by $\tilde{g}_{\mu\nu}$, i.e. $g_{\mu\nu} = a^2 \tilde{g}_{\mu\nu}$. The null geodesics remain the same; the affine parameters and the tangent vectors in the rescaled metric, on the other hand, are respectively given by $d\tilde{\lambda}=a^{-2}d\lambda$ and $\tilde{P}_{\mu}=a^2 P^{\mu}$ \cite{Wald:1984rg}. One also finds $P_{\mu}=\tilde{P}_{\mu}$. Using these relations the geodesic equation reads,
\begin{align}
    \label{eq:P0_geodesic}
   \frac{d\tilde{P}_0}{d\eta} =  -\frac{2}{5}\tilde{P}^0 \left(\frac{a}{a'}\right)'\partial_i\zeta n^i\,,
\end{align}
where the prime denotes a partial derivative with respect to conformal time. Using $P_0 = \tilde{P_0}$ and re-writing the above expression using the quantities defined in the original metric $g_{\mu\nu}$, one obtains at linear order in $\zeta$,
\begin{align}
    \frac{dP_0}{d\eta} & = -\frac{1}{5}a^2 P^0\,\partial_i\zeta n^i \nonumber\\
                        & = -\frac{1}{5}q\,\partial_i\zeta n^i\,.
\end{align} 
The Boltzmann equation then becomes 
\begin{align}
    \frac{df}{d\eta} = \del{f}{\eta} + \del{f}{x^i}n^i + \frac{1}{5}q\partial_i\zeta n^i \del{f}{q}
\end{align}
(Eq.~\eqref{eq:Boltzmann_zeta_2} of the main text).
Upon expanding the distribution function into a zeroth order term plus perturbations,
\begin{align}
    f(\eta,\vec x,q,\hn) = \bar{f}(q) - q\del{\bar f}{q}\Gamma(\eta,\vec x,q,\hn)\,,
\end{align}
the Boltzmann equation at first order reads,
\begin{align}
    \del{\Gamma}{\eta} + n^i\del{\Gamma}{x^i} = \frac{1}{5}\del{\zeta}{x^i}n^i\,.
\end{align}
In Fourier space, one finds
\begin{align}
     \del{\Gamma_k}{\eta}+ik\mu\Gamma_k = \frac{ik\mu}{5}\zeta_k\,,
\end{align}
with $\mu\equiv \hat k\cdot \hn$. This can be solved to give
\begin{align}
    \Gamma_k(\eta) = e^{ik\mu(\eta_{\rm in}-\eta)}\left[\Gamma_k(\eta_{\rm in},q,\hn) - \frac{1}{5}\zeta_k(\eta_{\rm in})\right]\,,
\end{align}
which is Eq.~\eqref{eq:Gamma_zeta} in Fourier space. 

We also provide here the original derivation of ref.~\cite{Alba:2015cms}. Starting from Eq.~\eqref{eq:P0_geodesic} and using the fact that the derivative along the geodesic can be written as 
\begin{align}
    \frac{d}{d\eta} = \del{}{\eta} + n^i\partial_i\,,
\end{align}
one obtains \cite{Alba:2017clw}
\begin{align}
    \frac{dP_0}{d\eta} =  -\frac{2}{5}P^0 \left(\frac{a}{a'}\right)'\frac{d}{d\eta}\zeta\,.
\end{align}
At zeroth order $P_0 = - P^0$ and one finds, 
\begin{align}
\label{two-eq}
    P_0(\eta) = P_0(\etai)\left[1+\frac{1}{5}\left(\zeta(\eta,\vx)-\zeta(\etai,\vx_i)\right)\right]\,,
\end{align}
where $\vec{x}_0 = \vec{x}_i + (\eta_0-\eta_i)\hn$. 
Equivalently, we can write for the comoving momentum $q$,
\begin{align}
       \frac{q}{q_i}= \left[1+\frac{1}{5}\left(\zeta(\eta,\vec x)-\zeta(\etai,\vec{x}_i)\right)\right].
       \label{eq:com_redshift_zgauge}
\end{align}
In this case one obtains a SGWB anisotropy equal to: 
\begin{align}
    \deltagw(q,\hn) &= \frac{1}{5}\zeta(\eta_i,\vec{x}_i)\del{\ln \bar{f}(q)}{\ln q}
\end{align}
(Eq.~\eqref{eq:deltagw_Alba} of the main text). Note the implicit assumption here that there is no initial perturbation to the distribution function, i.e. $f(\eta_i,\vec{x}_i) = \bar f(\eta_i)$, which is valid for adiabatic primordial perturbations when working in the uniform matter density gauge (see the discussion in Sec.~\ref{sec:Boltzmann_zeta}.).
The curvature perturbation at the observer's position can also be ignored since it only contributes to the monopole. Ref.~\cite{Alba:2015cms} considers the case of GW from single-field slow-roll inflation, with a distribution function $\bar{f}(q) \propto (q_0/q)^{2(\nu+1+\epsilon)}$.\footnote{Note that the quantity $\langle N \rangle$ of \cite{Alba:2015cms} is the same as $\bar{f}$ up to some constant factors.} Here $q_0$ is a reference frequency, $\epsilon$ the standard slow-roll parameter during inflation and $\nu \equiv 2/(1+3 \mathsf{w})$ with $\mathsf{w}$ evaluated at the time when the GW mode re-enters the horizon. Evaluating the SGWB anisotropy in terms of these quantities one arrives at \cite{Alba:2015cms}, 
\begin{align}
    \deltagw(q,\hn)=-\frac{2}{5}(\nu+1+\epsilon)\zeta(\eta_i,\vec{x}_i)\,.
\end{align}

\bibliographystyle{JHEP}
\bibliography{ref}

\providecommand{\href}[2]{#2}\begingroup\raggedright\begin{thebibliography}{100}

\bibitem{LIGOScientific:2019vic}
{\scshape LIGO Scientific, Virgo} collaboration, B.~P. Abbott et~al.,
  \emph{{Search for the isotropic stochastic background using data from
  Advanced LIGO\textquoteright{}s second observing run}},
  \href{http://dx.doi.org/10.1103/PhysRevD.100.061101}{\emph{Phys. Rev. D} {\bf
  100} (2019) 061101}, [\href{http://arxiv.org/abs/1903.02886}{{\tt
  1903.02886}}].

\bibitem{KAGRA:2021kbb}
{\scshape KAGRA, Virgo, LIGO Scientific} collaboration, R.~Abbott et~al.,
  \emph{{Upper limits on the isotropic gravitational-wave background from
  Advanced LIGO and Advanced Virgo\textquoteright{}s third observing run}},
  \href{http://dx.doi.org/10.1103/PhysRevD.104.022004}{\emph{Phys. Rev. D} {\bf
  104} (2021) 022004}, [\href{http://arxiv.org/abs/2101.12130}{{\tt
  2101.12130}}].

\bibitem{NANOGrav:2020bcs}
{\scshape NANOGrav} collaboration, Z.~Arzoumanian et~al., \emph{{The NANOGrav
  12.5 yr Data Set: Search for an Isotropic Stochastic Gravitational-wave
  Background}},
  \href{http://dx.doi.org/10.3847/2041-8213/abd401}{\emph{Astrophys. J. Lett.}
  {\bf 905} (2020) L34}, [\href{http://arxiv.org/abs/2009.04496}{{\tt
  2009.04496}}].

\bibitem{Regimbau:2011rp}
T.~Regimbau, \emph{{The astrophysical gravitational wave stochastic
  background}}, \href{http://dx.doi.org/10.1088/1674-4527/11/4/001}{\emph{Res.
  Astron. Astrophys.} {\bf 11} (2011) 369--390},
  [\href{http://arxiv.org/abs/1101.2762}{{\tt 1101.2762}}].

\bibitem{Caprini:2018mtu}
C.~Caprini and D.~G. Figueroa, \emph{{Cosmological Backgrounds of Gravitational
  Waves}}, \href{http://dx.doi.org/10.1088/1361-6382/aac608}{\emph{Class.
  Quant. Grav.} {\bf 35} (2018) 163001},
  [\href{http://arxiv.org/abs/1801.04268}{{\tt 1801.04268}}].

\bibitem{Guzzetti:2016mkm}
M.~Guzzetti, N.~Bartolo, M.~Liguori and S.~Matarrese, \emph{{Gravitational
  waves from inflation}},
  \href{http://dx.doi.org/10.1393/ncr/i2016-10127-1}{\emph{Riv. Nuovo Cim.}
  {\bf 39} (2016) 399--495}, [\href{http://arxiv.org/abs/1605.01615}{{\tt
  1605.01615}}].

\bibitem{Punturo:2010zz}
M.~Punturo et~al., \emph{{The Einstein Telescope: A third-generation
  gravitational wave observatory}},
  \href{http://dx.doi.org/10.1088/0264-9381/27/19/194002}{\emph{Class. Quant.
  Grav.} {\bf 27} (2010) 194002}.

\bibitem{Reitze:2019iox}
D.~Reitze et~al., \emph{{Cosmic Explorer: The U.S. Contribution to
  Gravitational-Wave Astronomy beyond LIGO}}, {\emph{Bull. Am. Astron. Soc.}
  {\bf 51} (2019) 035}, [\href{http://arxiv.org/abs/1907.04833}{{\tt
  1907.04833}}].

\bibitem{amaroseoane2017laser}
P.~Amaro-Seoane, H.~Audley, S.~Babak, J.~Baker, E.~Barausse, P.~Bender et~al.,
  \emph{Laser interferometer space antenna},
  \href{http://arxiv.org/abs/1702.00786}{{\tt 1702.00786}}.

\bibitem{Hu:2017mde}
W.-R. Hu and Y.-L. Wu, \emph{{The Taiji Program in Space for gravitational wave
  physics and the nature of gravity}},
  \href{http://dx.doi.org/10.1093/nsr/nwx116}{\emph{Natl. Sci. Rev.} {\bf 4}
  (2017) 685--686}.

\bibitem{Giare:2020plo}
W.~Giar\`e, F.~Renzi and A.~Melchiorri, \emph{{Higher-Curvature Corrections and
  Tensor Modes}},
  \href{http://dx.doi.org/10.1103/PhysRevD.103.043515}{\emph{Phys. Rev. D} {\bf
  103} (2021) 043515}, [\href{http://arxiv.org/abs/2012.00527}{{\tt
  2012.00527}}].

\bibitem{Bartolo:2017szm}
N.~Bartolo and G.~Orlando, \emph{{Parity breaking signatures from a
  Chern-Simons coupling during inflation: the case of non-Gaussian
  gravitational waves}},
  \href{http://dx.doi.org/10.1088/1475-7516/2017/07/034}{\emph{JCAP} {\bf 07}
  (2017) 034}, [\href{http://arxiv.org/abs/1706.04627}{{\tt 1706.04627}}].

\bibitem{Burrage:2010cu}
C.~Burrage, C.~de~Rham, D.~Seery and A.~J. Tolley, \emph{{Galileon inflation}},
  \href{http://dx.doi.org/10.1088/1475-7516/2011/01/014}{\emph{JCAP} {\bf 01}
  (2011) 014}, [\href{http://arxiv.org/abs/1009.2497}{{\tt 1009.2497}}].

\bibitem{Tolley:2009fg}
A.~J. Tolley and M.~Wyman, \emph{{The Gelaton Scenario: Equilateral
  non-Gaussianity from multi-field dynamics}},
  \href{http://dx.doi.org/10.1103/PhysRevD.81.043502}{\emph{Phys. Rev. D} {\bf
  81} (2010) 043502}, [\href{http://arxiv.org/abs/0910.1853}{{\tt 0910.1853}}].

\bibitem{Ade:2015lrj}
{\scshape Planck} collaboration, P.~A.~R. Ade et~al., \emph{{Planck 2015
  results. XX. Constraints on inflation}},
  \href{http://arxiv.org/abs/1502.02114}{{\tt 1502.02114}}.

\bibitem{Akrami:2018odb}
{\scshape Planck} collaboration, Y.~Akrami et~al., \emph{{Planck 2018 results.
  X. Constraints on inflation}},  \href{http://arxiv.org/abs/1807.06211}{{\tt
  1807.06211}}.

\bibitem{BICEP:2021xfz}
{\scshape BICEP, Keck} collaboration, P.~A.~R. Ade et~al., \emph{{Improved
  Constraints on Primordial Gravitational Waves using Planck, WMAP, and
  BICEP/Keck Observations through the 2018 Observing Season}},
  \href{http://dx.doi.org/10.1103/PhysRevLett.127.151301}{\emph{Phys. Rev.
  Lett.} {\bf 127} (2021) 151301}, [\href{http://arxiv.org/abs/2110.00483}{{\tt
  2110.00483}}].

\bibitem{Tristram:2021tvh}
M.~Tristram et~al., \emph{{Improved limits on the tensor-to-scalar ratio using
  BICEP and Planck}},  \href{http://arxiv.org/abs/2112.07961}{{\tt
  2112.07961}}.

\bibitem{CMB-S4:2020lpa}
{\scshape CMB-S4} collaboration, K.~Abazajian et~al., \emph{{CMB-S4:
  Forecasting Constraints on Primordial Gravitational Waves}},
  \href{http://arxiv.org/abs/2008.12619}{{\tt 2008.12619}}.

\bibitem{SimonsObservatory:2018koc}
{\scshape Simons Observatory} collaboration, P.~Ade et~al., \emph{{The Simons
  Observatory: Science goals and forecasts}},
  \href{http://dx.doi.org/10.1088/1475-7516/2019/02/056}{\emph{JCAP} {\bf 02}
  (2019) 056}, [\href{http://arxiv.org/abs/1808.07445}{{\tt 1808.07445}}].

\bibitem{LiteBIRD:2022cnt}
{\scshape LiteBIRD} collaboration, E.~Allys et~al., \emph{{Probing Cosmic
  Inflation with the LiteBIRD Cosmic Microwave Background Polarization
  Survey}},  \href{http://arxiv.org/abs/2202.02773}{{\tt 2202.02773}}.

\bibitem{Starobinsky:1980te}
A.~A. Starobinsky, \emph{{A New Type of Isotropic Cosmological Models Without
  Singularity}},
  \href{http://dx.doi.org/10.1016/0370-2693(80)90670-X}{\emph{Phys. Lett.} {\bf
  B91} (1980) 99--102}.

\bibitem{Kehagias:2013mya}
A.~Kehagias, A.~Moradinezhad~Dizgah and A.~Riotto, \emph{{Remarks on the
  Starobinsky model of inflation and its descendants}},
  \href{http://dx.doi.org/10.1103/PhysRevD.89.043527}{\emph{Phys. Rev. D} {\bf
  89} (2014) 043527}, [\href{http://arxiv.org/abs/1312.1155}{{\tt 1312.1155}}].

\bibitem{Crowder:2005nr}
J.~Crowder and N.~J. Cornish, \emph{{Beyond LISA: Exploring future
  gravitational wave missions}},
  \href{http://dx.doi.org/10.1103/PhysRevD.72.083005}{\emph{Phys. Rev. D} {\bf
  72} (2005) 083005}, [\href{http://arxiv.org/abs/gr-qc/0506015}{{\tt
  gr-qc/0506015}}].

\bibitem{Cook:2011hg}
J.~L. Cook and L.~Sorbo, \emph{{Particle production during inflation and
  gravitational waves detectable by ground-based interferometers}},
  \href{http://dx.doi.org/10.1103/PhysRevD.85.023534}{\emph{Phys. Rev. D} {\bf
  85} (2012) 023534}, [\href{http://arxiv.org/abs/1109.0022}{{\tt 1109.0022}}].

\bibitem{Barnaby:2011qe}
N.~Barnaby, E.~Pajer and M.~Peloso, \emph{{Gauge Field Production in Axion
  Inflation: Consequences for Monodromy, non-Gaussianity in the CMB, and
  Gravitational Waves at Interferometers}},
  \href{http://dx.doi.org/10.1103/PhysRevD.85.023525}{\emph{Phys. Rev. D} {\bf
  85} (2012) 023525}, [\href{http://arxiv.org/abs/1110.3327}{{\tt 1110.3327}}].

\bibitem{Barnaby:2012xt}
N.~Barnaby, J.~Moxon, R.~Namba, M.~Peloso, G.~Shiu and P.~Zhou, \emph{{Gravity
  waves and non-Gaussian features from particle production in a sector
  gravitationally coupled to the inflaton}},
  \href{http://dx.doi.org/10.1103/PhysRevD.86.103508}{\emph{Phys. Rev. D} {\bf
  86} (2012) 103508}, [\href{http://arxiv.org/abs/1206.6117}{{\tt 1206.6117}}].

\bibitem{Dimastrogiovanni:2016fuu}
E.~Dimastrogiovanni, M.~Fasiello and T.~Fujita, \emph{{Primordial Gravitational
  Waves from Axion-Gauge Fields Dynamics}},
  \href{http://dx.doi.org/10.1088/1475-7516/2017/01/019}{\emph{JCAP} {\bf 01}
  (2017) 019}, [\href{http://arxiv.org/abs/1608.04216}{{\tt 1608.04216}}].

\bibitem{Garcia-Bellido:2016dkw}
J.~Garcia-Bellido, M.~Peloso and C.~Unal, \emph{{Gravitational waves at
  interferometer scales and primordial black holes in axion inflation}},
  \href{http://dx.doi.org/10.1088/1475-7516/2016/12/031}{\emph{JCAP} {\bf 12}
  (2016) 031}, [\href{http://arxiv.org/abs/1610.03763}{{\tt 1610.03763}}].

\bibitem{Thorne:2017jft}
B.~Thorne, T.~Fujita, M.~Hazumi, N.~Katayama, E.~Komatsu and M.~Shiraishi,
  \emph{{Finding the chiral gravitational wave background of an axion-SU(2)
  inflationary model using CMB observations and laser interferometers}},
  \href{http://dx.doi.org/10.1103/PhysRevD.97.043506}{\emph{Phys. Rev. D} {\bf
  97} (2018) 043506}, [\href{http://arxiv.org/abs/1707.03240}{{\tt
  1707.03240}}].

\bibitem{Domcke:2018eki}
V.~Domcke and K.~Mukaida, \emph{{Gauge Field and Fermion Production during
  Axion Inflation}},
  \href{http://dx.doi.org/10.1088/1475-7516/2018/11/020}{\emph{JCAP} {\bf 11}
  (2018) 020}, [\href{http://arxiv.org/abs/1806.08769}{{\tt 1806.08769}}].

\bibitem{Bordin:2018pca}
L.~Bordin, P.~Creminelli, A.~Khmelnitsky and L.~Senatore, \emph{{Light
  Particles with Spin in Inflation}},
  \href{http://dx.doi.org/10.1088/1475-7516/2018/10/013}{\emph{JCAP} {\bf 10}
  (2018) 013}, [\href{http://arxiv.org/abs/1806.10587}{{\tt 1806.10587}}].

\bibitem{Iacconi:2019vgc}
L.~Iacconi, M.~Fasiello, H.~Assadullahi, E.~Dimastrogiovanni and D.~Wands,
  \emph{{Interferometer Constraints on the Inflationary Field Content}},
  \href{http://dx.doi.org/10.1088/1475-7516/2020/03/031}{\emph{JCAP} {\bf 03}
  (2020) 031}, [\href{http://arxiv.org/abs/1910.12921}{{\tt 1910.12921}}].

\bibitem{Campeti:2020xwn}
P.~Campeti, E.~Komatsu, D.~Poletti and C.~Baccigalupi, \emph{{Measuring the
  spectrum of primordial gravitational waves with CMB, PTA and Laser
  Interferometers}},
  \href{http://dx.doi.org/10.1088/1475-7516/2021/01/012}{\emph{JCAP} {\bf 01}
  (2021) 012}, [\href{http://arxiv.org/abs/2007.04241}{{\tt 2007.04241}}].

\bibitem{Iacconi:2020yxn}
L.~Iacconi, M.~Fasiello, H.~Assadullahi and D.~Wands, \emph{{Small-scale Tests
  of Inflation}},
  \href{http://dx.doi.org/10.1088/1475-7516/2020/12/005}{\emph{JCAP} {\bf 12}
  (2020) 005}, [\href{http://arxiv.org/abs/2008.00452}{{\tt 2008.00452}}].

\bibitem{Endlich:2012pz}
S.~Endlich, A.~Nicolis and J.~Wang, \emph{{Solid Inflation}},
  \href{http://dx.doi.org/10.1088/1475-7516/2013/10/011}{\emph{JCAP} {\bf 1310}
  (2013) 011}, [\href{http://arxiv.org/abs/1210.0569}{{\tt 1210.0569}}].

\bibitem{Bartolo:2015qvr}
N.~Bartolo, D.~Cannone, A.~Ricciardone and G.~Tasinato, \emph{{Distinctive
  signatures of space-time diffeomorphism breaking in EFT of inflation}},
  \href{http://dx.doi.org/10.1088/1475-7516/2016/03/044}{\emph{JCAP} {\bf 03}
  (2016) 044}, [\href{http://arxiv.org/abs/1511.07414}{{\tt 1511.07414}}].

\bibitem{Ricciardone:2016lym}
A.~Ricciardone and G.~Tasinato, \emph{{Primordial gravitational waves in
  supersolid inflation}},
  \href{http://dx.doi.org/10.1103/PhysRevD.96.023508}{\emph{Phys. Rev. D} {\bf
  96} (2017) 023508}, [\href{http://arxiv.org/abs/1611.04516}{{\tt
  1611.04516}}].

\bibitem{Celoria:2020diz}
M.~Celoria, D.~Comelli, L.~Pilo and R.~Rollo, \emph{{Boosting GWs in Supersolid
  Inflation}}, \href{http://dx.doi.org/10.1007/JHEP01(2021)185}{\emph{JHEP}
  {\bf 01} (2021) 185}, [\href{http://arxiv.org/abs/2010.02023}{{\tt
  2010.02023}}].

\bibitem{Celoria:2021cxq}
M.~Celoria, D.~Comelli, L.~Pilo and R.~Rollo, \emph{{Primordial Non-Gaussianity
  in Supersolid Inflation}},  \href{http://arxiv.org/abs/2103.10402}{{\tt
  2103.10402}}.

\bibitem{Mylova:2018yap}
M.~Mylova, O.~\"Ozsoy, S.~Parameswaran, G.~Tasinato and I.~Zavala, \emph{{A new
  mechanism to enhance primordial tensor fluctuations in single field
  inflation}},
  \href{http://dx.doi.org/10.1088/1475-7516/2018/12/024}{\emph{JCAP} {\bf 12}
  (2018) 024}, [\href{http://arxiv.org/abs/1808.10475}{{\tt 1808.10475}}].

\bibitem{Ozsoy:2019slf}
O.~Ozsoy, M.~Mylova, S.~Parameswaran, C.~Powell, G.~Tasinato and I.~Zavala,
  \emph{{Squeezed tensor non-Gaussianity in non-attractor inflation}},
  \href{http://dx.doi.org/10.1088/1475-7516/2019/09/036}{\emph{JCAP} {\bf 09}
  (2019) 036}, [\href{http://arxiv.org/abs/1902.04976}{{\tt 1902.04976}}].

\bibitem{Dimastrogiovanni:2019bfl}
E.~Dimastrogiovanni, M.~Fasiello and G.~Tasinato, \emph{{Searching for Fossil
  Fields in the Gravity Sector}},
  \href{http://dx.doi.org/10.1103/PhysRevLett.124.061302}{\emph{Phys. Rev.
  Lett.} {\bf 124} (2020) 061302}, [\href{http://arxiv.org/abs/1906.07204}{{\tt
  1906.07204}}].

\bibitem{Adshead:2020bji}
P.~Adshead, N.~Afshordi, E.~Dimastrogiovanni, M.~Fasiello, E.~A. Lim and
  G.~Tasinato, \emph{{Multimessenger Cosmology: correlating CMB and SGWB
  measurements}},
  \href{http://dx.doi.org/10.1103/PhysRevD.103.023532}{\emph{Phys. Rev. D} {\bf
  103} (2021) 023532}, [\href{http://arxiv.org/abs/2004.06619}{{\tt
  2004.06619}}].

\bibitem{Malhotra:2020ket}
A.~Malhotra, E.~Dimastrogiovanni, M.~Fasiello and M.~Shiraishi,
  \emph{{Cross-correlations as a Diagnostic Tool for Primordial Gravitational
  Waves}}, \href{http://dx.doi.org/10.1088/1475-7516/2021/03/088}{\emph{JCAP}
  {\bf 03} (2021) 088}, [\href{http://arxiv.org/abs/2012.03498}{{\tt
  2012.03498}}].

\bibitem{Dimastrogiovanni:2021mfs}
E.~Dimastrogiovanni, M.~Fasiello, A.~Malhotra, P.~D. Meerburg and G.~Orlando,
  \emph{{Testing the early universe with anisotropies of the gravitational wave
  background}},
  \href{http://dx.doi.org/10.1088/1475-7516/2022/02/040}{\emph{JCAP} {\bf 02}
  (2022) 040}, [\href{http://arxiv.org/abs/2109.03077}{{\tt 2109.03077}}].

\bibitem{Dimastrogiovanni:2022afr}
E.~Dimastrogiovanni, M.~Fasiello and L.~Pinol, \emph{{Primordial Stochastic
  Gravitational Wave Background Anisotropies: in-in Formalization and
  Applications}},  \href{http://arxiv.org/abs/2203.17192}{{\tt 2203.17192}}.

\bibitem{Alba:2015cms}
V.~Alba and J.~Maldacena, \emph{{Primordial gravity wave background
  anisotropies}}, \href{http://dx.doi.org/10.1007/JHEP03(2016)115}{\emph{JHEP}
  {\bf 03} (2016) 115}, [\href{http://arxiv.org/abs/1512.01531}{{\tt
  1512.01531}}].

\bibitem{Contaldi:2016koz}
C.~R. Contaldi, \emph{{Anisotropies of Gravitational Wave Backgrounds: A Line
  Of Sight Approach}},
  \href{http://dx.doi.org/10.1016/j.physletb.2017.05.020}{\emph{Phys. Lett. B}
  {\bf 771} (2017) 9--12}, [\href{http://arxiv.org/abs/1609.08168}{{\tt
  1609.08168}}].

\bibitem{Bartolo:2019oiq}
N.~Bartolo, D.~Bertacca, S.~Matarrese, M.~Peloso, A.~Ricciardone, A.~Riotto
  et~al., \emph{{Anisotropies and non-Gaussianity of the Cosmological
  Gravitational Wave Background}},
  \href{http://dx.doi.org/10.1103/PhysRevD.100.121501}{\emph{Phys. Rev. D} {\bf
  100} (2019) 121501}, [\href{http://arxiv.org/abs/1908.00527}{{\tt
  1908.00527}}].

\bibitem{Bartolo:2019yeu}
N.~Bartolo, D.~Bertacca, S.~Matarrese, M.~Peloso, A.~Ricciardone, A.~Riotto
  et~al., \emph{{Characterizing the cosmological gravitational wave background:
  Anisotropies and non-Gaussianity}},
  \href{http://dx.doi.org/10.1103/PhysRevD.102.023527}{\emph{Phys. Rev. D} {\bf
  102} (2020) 023527}, [\href{http://arxiv.org/abs/1912.09433}{{\tt
  1912.09433}}].

\bibitem{Domcke:2020xmn}
V.~Domcke, R.~Jinno and H.~Rubira, \emph{{Deformation of the gravitational wave
  spectrum by density perturbations}},
  \href{http://dx.doi.org/10.1088/1475-7516/2020/06/046}{\emph{JCAP} {\bf 06}
  (2020) 046}, [\href{http://arxiv.org/abs/2002.11083}{{\tt 2002.11083}}].

\bibitem{Geller:2018mwu}
M.~Geller, A.~Hook, R.~Sundrum and Y.~Tsai, \emph{{Primordial Anisotropies in
  the Gravitational Wave Background from Cosmological Phase Transitions}},
  \href{http://dx.doi.org/10.1103/PhysRevLett.121.201303}{\emph{Phys. Rev.
  Lett.} {\bf 121} (2018) 201303}, [\href{http://arxiv.org/abs/1803.10780}{{\tt
  1803.10780}}].

\bibitem{Kumar:2021ffi}
S.~Kumar, R.~Sundrum and Y.~Tsai, \emph{{Non-Gaussian stochastic gravitational
  waves from phase transitions}},
  \href{http://dx.doi.org/10.1007/JHEP11(2021)107}{\emph{JHEP} {\bf 11} (2021)
  107}, [\href{http://arxiv.org/abs/2102.05665}{{\tt 2102.05665}}].

\bibitem{Li:2021iva}
Y.~Li, F.~P. Huang, X.~Wang and X.~Zhang, \emph{{Anisotropy of phase transition
  gravitational wave}},  \href{http://arxiv.org/abs/2112.01409}{{\tt
  2112.01409}}.

\bibitem{Olmez:2011cg}
S.~Olmez, V.~Mandic and X.~Siemens, \emph{{Anisotropies in the
  Gravitational-Wave Stochastic Background}},
  \href{http://dx.doi.org/10.1088/1475-7516/2012/07/009}{\emph{JCAP} {\bf 07}
  (2012) 009}, [\href{http://arxiv.org/abs/1106.5555}{{\tt 1106.5555}}].

\bibitem{Kuroyanagi:2016ugi}
S.~Kuroyanagi, K.~Takahashi, N.~Yonemaru and H.~Kumamoto, \emph{{Anisotropies
  in the gravitational wave background as a probe of the cosmic string
  network}}, \href{http://dx.doi.org/10.1103/PhysRevD.95.043531}{\emph{Phys.
  Rev. D} {\bf 95} (2017) 043531}, [\href{http://arxiv.org/abs/1604.00332}{{\tt
  1604.00332}}].

\bibitem{Jenkins:2018nty}
A.~C. Jenkins and M.~Sakellariadou, \emph{{Anisotropies in the stochastic
  gravitational-wave background: Formalism and the cosmic string case}},
  \href{http://dx.doi.org/10.1103/PhysRevD.98.063509}{\emph{Phys. Rev. D} {\bf
  98} (2018) 063509}, [\href{http://arxiv.org/abs/1802.06046}{{\tt
  1802.06046}}].

\bibitem{Cai:2021dgx}
R.-G. Cai, Z.-K. Guo and J.~Liu, \emph{{A New Picture of Cosmic String
  Evolution and Anisotropic Stochastic Gravitational-Wave Background}},
  \href{http://arxiv.org/abs/2112.10131}{{\tt 2112.10131}}.

\bibitem{DallArmi:2020dar}
L.~Valbusa~Dall'Armi, A.~Ricciardone, N.~Bartolo, D.~Bertacca and S.~Matarrese,
  \emph{{Imprint of relativistic particles on the anisotropies of the
  stochastic gravitational-wave background}},
  \href{http://dx.doi.org/10.1103/PhysRevD.103.023522}{\emph{Phys. Rev. D} {\bf
  103} (2021) 023522}, [\href{http://arxiv.org/abs/2007.01215}{{\tt
  2007.01215}}].

\bibitem{Braglia:2021fxn}
M.~Braglia and S.~Kuroyanagi, \emph{{Probing prerecombination physics by the
  cross-correlation of stochastic gravitational waves and CMB anisotropies}},
  \href{http://dx.doi.org/10.1103/PhysRevD.104.123547}{\emph{Phys. Rev. D} {\bf
  104} (2021) 123547}, [\href{http://arxiv.org/abs/2106.03786}{{\tt
  2106.03786}}].

\bibitem{Ricciardone:2021kel}
A.~Ricciardone, L.~V. Dall'Armi, N.~Bartolo, D.~Bertacca, M.~Liguori and
  S.~Matarrese, \emph{{Cross-Correlating Astrophysical and Cosmological
  Gravitational Wave Backgrounds with the Cosmic Microwave Background}},
  \href{http://dx.doi.org/10.1103/PhysRevLett.127.271301}{\emph{Phys. Rev.
  Lett.} {\bf 127} (2021) 271301}, [\href{http://arxiv.org/abs/2106.02591}{{\tt
  2106.02591}}].

\bibitem{Cusin:2017fwz}
G.~Cusin, C.~Pitrou and J.-P. Uzan, \emph{{Anisotropy of the astrophysical
  gravitational wave background: Analytic expression of the angular power
  spectrum and correlation with cosmological observations}},
  \href{http://dx.doi.org/10.1103/PhysRevD.96.103019}{\emph{Phys. Rev. D} {\bf
  96} (2017) 103019}, [\href{http://arxiv.org/abs/1704.06184}{{\tt
  1704.06184}}].

\bibitem{Cusin:2018rsq}
G.~Cusin, I.~Dvorkin, C.~Pitrou and J.-P. Uzan, \emph{{First predictions of the
  angular power spectrum of the astrophysical gravitational wave background}},
  \href{http://dx.doi.org/10.1103/PhysRevLett.120.231101}{\emph{Phys. Rev.
  Lett.} {\bf 120} (2018) 231101}, [\href{http://arxiv.org/abs/1803.03236}{{\tt
  1803.03236}}].

\bibitem{Jenkins:2018uac}
A.~C. Jenkins, M.~Sakellariadou, T.~Regimbau and E.~Slezak, \emph{{Anisotropies
  in the astrophysical gravitational-wave background: Predictions for the
  detection of compact binaries by LIGO and Virgo}},
  \href{http://dx.doi.org/10.1103/PhysRevD.98.063501}{\emph{Phys. Rev. D} {\bf
  98} (2018) 063501}, [\href{http://arxiv.org/abs/1806.01718}{{\tt
  1806.01718}}].

\bibitem{Jenkins:2018kxc}
A.~C. Jenkins, R.~O'Shaughnessy, M.~Sakellariadou and D.~Wysocki,
  \emph{{Anisotropies in the astrophysical gravitational-wave background: The
  impact of black hole distributions}},
  \href{http://dx.doi.org/10.1103/PhysRevLett.122.111101}{\emph{Phys. Rev.
  Lett.} {\bf 122} (2019) 111101}, [\href{http://arxiv.org/abs/1810.13435}{{\tt
  1810.13435}}].

\bibitem{Jenkins:2019uzp}
A.~C. Jenkins and M.~Sakellariadou, \emph{{Shot noise in the astrophysical
  gravitational-wave background}},
  \href{http://dx.doi.org/10.1103/PhysRevD.100.063508}{\emph{Phys. Rev. D} {\bf
  100} (2019) 063508}, [\href{http://arxiv.org/abs/1902.07719}{{\tt
  1902.07719}}].

\bibitem{Cusin:2019jhg}
G.~Cusin, I.~Dvorkin, C.~Pitrou and J.-P. Uzan, \emph{{Stochastic gravitational
  wave background anisotropies in the mHz band: astrophysical dependencies}},
  \href{http://dx.doi.org/10.1093/mnrasl/slz182}{\emph{Mon. Not. Roy. Astron.
  Soc.} {\bf 493} (2020) L1--L5}, [\href{http://arxiv.org/abs/1904.07757}{{\tt
  1904.07757}}].

\bibitem{Cusin:2019jpv}
G.~Cusin, I.~Dvorkin, C.~Pitrou and J.-P. Uzan, \emph{{Properties of the
  stochastic astrophysical gravitational wave background: astrophysical sources
  dependencies}},
  \href{http://dx.doi.org/10.1103/PhysRevD.100.063004}{\emph{Phys. Rev. D} {\bf
  100} (2019) 063004}, [\href{http://arxiv.org/abs/1904.07797}{{\tt
  1904.07797}}].

\bibitem{Bertacca:2019fnt}
D.~Bertacca, A.~Ricciardone, N.~Bellomo, A.~C. Jenkins, S.~Matarrese,
  A.~Raccanelli et~al., \emph{{Projection effects on the observed angular
  spectrum of the astrophysical stochastic gravitational wave background}},
  \href{http://dx.doi.org/10.1103/PhysRevD.101.103513}{\emph{Phys. Rev. D} {\bf
  101} (2020) 103513}, [\href{http://arxiv.org/abs/1909.11627}{{\tt
  1909.11627}}].

\bibitem{Pitrou:2019rjz}
C.~Pitrou, G.~Cusin and J.-P. Uzan, \emph{{Unified view of anisotropies in the
  astrophysical gravitational-wave background}},
  \href{http://dx.doi.org/10.1103/PhysRevD.101.081301}{\emph{Phys. Rev. D} {\bf
  101} (2020) 081301}, [\href{http://arxiv.org/abs/1910.04645}{{\tt
  1910.04645}}].

\bibitem{Capurri:2021zli}
G.~Capurri, A.~Lapi, C.~Baccigalupi, L.~Boco, G.~Scelfo and T.~Ronconi,
  \emph{{Intensity and anisotropies of the stochastic Gravitational Wave
  background from merging compact binaries in galaxies}},
  \href{http://arxiv.org/abs/2103.12037}{{\tt 2103.12037}}.

\bibitem{Bellomo:2021mer}
N.~Bellomo, D.~Bertacca, A.~C. Jenkins, S.~Matarrese, A.~Raccanelli,
  T.~Regimbau et~al., \emph{{CLASS\_GWB: robust modeling of the astrophysical
  gravitational wave background anisotropies}},
  \href{http://arxiv.org/abs/2110.15059}{{\tt 2110.15059}}.

\bibitem{Matarrese:1997ay}
S.~Matarrese, S.~Mollerach and M.~Bruni, \emph{{Second order perturbations of
  the Einstein-de Sitter universe}},
  \href{http://dx.doi.org/10.1103/PhysRevD.58.043504}{\emph{Phys. Rev. D} {\bf
  58} (1998) 043504}, [\href{http://arxiv.org/abs/astro-ph/9707278}{{\tt
  astro-ph/9707278}}].

\bibitem{Ananda:2006af}
K.~N. Ananda, C.~Clarkson and D.~Wands, \emph{{The Cosmological gravitational
  wave background from primordial density perturbations}},
  \href{http://dx.doi.org/10.1103/PhysRevD.75.123518}{\emph{Phys. Rev. D} {\bf
  75} (2007) 123518}, [\href{http://arxiv.org/abs/gr-qc/0612013}{{\tt
  gr-qc/0612013}}].

\bibitem{Baumann:2007zm}
D.~Baumann, P.~J. Steinhardt, K.~Takahashi and K.~Ichiki, \emph{{Gravitational
  Wave Spectrum Induced by Primordial Scalar Perturbations}},
  \href{http://dx.doi.org/10.1103/PhysRevD.76.084019}{\emph{Phys. Rev. D} {\bf
  76} (2007) 084019}, [\href{http://arxiv.org/abs/hep-th/0703290}{{\tt
  hep-th/0703290}}].

\bibitem{Domenech:2021ztg}
G.~Dom\`enech, \emph{{Scalar Induced Gravitational Waves Review}},
  \href{http://dx.doi.org/10.3390/universe7110398}{\emph{Universe} {\bf 7}
  (2021) 398}, [\href{http://arxiv.org/abs/2109.01398}{{\tt 2109.01398}}].

\bibitem{Sasaki:2018dmp}
M.~Sasaki, T.~Suyama, T.~Tanaka and S.~Yokoyama, \emph{{Primordial black
  holes\textemdash{}perspectives in gravitational wave astronomy}},
  \href{http://dx.doi.org/10.1088/1361-6382/aaa7b4}{\emph{Class. Quant. Grav.}
  {\bf 35} (2018) 063001}, [\href{http://arxiv.org/abs/1801.05235}{{\tt
  1801.05235}}].

\bibitem{Auclair:2022lcg}
P.~Auclair et~al., \emph{{Cosmology with the Laser Interferometer Space
  Antenna}},  \href{http://arxiv.org/abs/2204.05434}{{\tt 2204.05434}}.

\bibitem{Isaacson_geomoptics}
R.~A. Isaacson, \emph{Gravitational radiation in the limit of high frequency.
  i. the linear approximation and geometrical optics},
  \href{http://dx.doi.org/10.1103/PhysRev.166.1263}{\emph{Phys. Rev.} {\bf 166}
  (Feb, 1968) 1263--1271}.

\bibitem{Misner:1973prb}
C.~W. Misner, K.~S. Thorne and J.~A. Wheeler, \emph{{Gravitation}}.
\newblock W. H. Freeman, San Francisco, 1973.

\bibitem{Maggiore:1999vm}
M.~Maggiore, \emph{{Gravitational wave experiments and early universe
  cosmology}},
  \href{http://dx.doi.org/10.1016/S0370-1573(99)00102-7}{\emph{Phys. Rept.}
  {\bf 331} (2000) 283--367}, [\href{http://arxiv.org/abs/gr-qc/9909001}{{\tt
  gr-qc/9909001}}].

\bibitem{Abramo:1997hu}
L.~R.~W. Abramo, R.~H. Brandenberger and V.~F. Mukhanov, \emph{{The Energy -
  momentum tensor for cosmological perturbations}},
  \href{http://dx.doi.org/10.1103/PhysRevD.56.3248}{\emph{Phys. Rev. D} {\bf
  56} (1997) 3248--3257}, [\href{http://arxiv.org/abs/gr-qc/9704037}{{\tt
  gr-qc/9704037}}].

\bibitem{Dodelson:2003ft}
S.~Dodelson, \emph{{Modern Cosmology}}.
\newblock Academic Press, Amsterdam, 2003.

\bibitem{Creminelli:2011sq}
P.~Creminelli, C.~Pitrou and F.~Vernizzi, \emph{{The CMB bispectrum in the
  squeezed limit}},
  \href{http://dx.doi.org/10.1088/1475-7516/2011/11/025}{\emph{JCAP} {\bf 11}
  (2011) 025}, [\href{http://arxiv.org/abs/1109.1822}{{\tt 1109.1822}}].

\bibitem{Hwang:2001hr}
J.~Hwang, T.~Padmanabhan, O.~Lahav and H.~Noh, \emph{{On the 1/3 factor in the
  CMB Sachs-Wolfe effect}},
  \href{http://dx.doi.org/10.1103/PhysRevD.65.043005}{\emph{Phys. Rev. D} {\bf
  65} (2002) 043005}, [\href{http://arxiv.org/abs/astro-ph/0107307}{{\tt
  astro-ph/0107307}}].

\bibitem{Bodas:2022zca}
A.~Bodas and R.~Sundrum, \emph{{Primordial Clocks within Stochastic
  Gravitational Wave Anisotropies}},
  \href{http://arxiv.org/abs/2205.04482}{{\tt 2205.04482}}.

\bibitem{Boubekeur:2008kn}
L.~Boubekeur, P.~Creminelli, J.~Norena and F.~Vernizzi, \emph{{Action approach
  to cosmological perturbations: the 2nd order metric in matter dominance}},
  \href{http://dx.doi.org/10.1088/1475-7516/2008/08/028}{\emph{JCAP} {\bf 08}
  (2008) 028}, [\href{http://arxiv.org/abs/0806.1016}{{\tt 0806.1016}}].

\bibitem{Pi:2020otn}
S.~Pi and M.~Sasaki, \emph{{Gravitational Waves Induced by Scalar Perturbations
  with a Lognormal Peak}},
  \href{http://dx.doi.org/10.1088/1475-7516/2020/09/037}{\emph{JCAP} {\bf 09}
  (2020) 037}, [\href{http://arxiv.org/abs/2005.12306}{{\tt 2005.12306}}].

\bibitem{Bartolo:2019zvb}
N.~Bartolo, D.~Bertacca, V.~De~Luca, G.~Franciolini, S.~Matarrese, M.~Peloso
  et~al., \emph{{Gravitational wave anisotropies from primordial black holes}},
  \href{http://dx.doi.org/10.1088/1475-7516/2020/02/028}{\emph{JCAP} {\bf 02}
  (2020) 028}, [\href{http://arxiv.org/abs/1909.12619}{{\tt 1909.12619}}].

\bibitem{Saito:2010pbh}
R.~{Saito} and J.~{Yokoyama}, \emph{{Gravitational-Wave Constraints on the
  Abundance of Primordial Black Holes}},
  \href{http://dx.doi.org/10.1143/PTP.123.867}{\emph{Progress of Theoretical
  Physics} {\bf 123} (May, 2010) 867--886},
  [\href{http://arxiv.org/abs/0912.5317}{{\tt 0912.5317}}].

\bibitem{Byrnes:2018txb}
C.~T. Byrnes, P.~S. Cole and S.~P. Patil, \emph{{Steepest growth of the power
  spectrum and primordial black holes}},
  \href{http://dx.doi.org/10.1088/1475-7516/2019/06/028}{\emph{JCAP} {\bf 06}
  (2019) 028}, [\href{http://arxiv.org/abs/1811.11158}{{\tt 1811.11158}}].

\bibitem{Carrilho:2019oqg}
P.~Carrilho, K.~A. Malik and D.~J. Mulryne, \emph{{Dissecting the growth of the
  power spectrum for primordial black holes}},
  \href{http://dx.doi.org/10.1103/PhysRevD.100.103529}{\emph{Phys. Rev. D} {\bf
  100} (2019) 103529}, [\href{http://arxiv.org/abs/1907.05237}{{\tt
  1907.05237}}].

\bibitem{Ozsoy:2019lyy}
O.~\"Ozsoy and G.~Tasinato, \emph{{On the slope of the curvature power spectrum
  in non-attractor inflation}},
  \href{http://dx.doi.org/10.1088/1475-7516/2020/04/048}{\emph{JCAP} {\bf 04}
  (2020) 048}, [\href{http://arxiv.org/abs/1912.01061}{{\tt 1912.01061}}].

\bibitem{Tasinato:2020vdk}
G.~Tasinato, \emph{{An analytic approach to non-slow-roll inflation}},
  \href{http://dx.doi.org/10.1103/PhysRevD.103.023535}{\emph{Phys. Rev. D} {\bf
  103} (2021) 023535}, [\href{http://arxiv.org/abs/2012.02518}{{\tt
  2012.02518}}].

\bibitem{Pi:2017gih}
S.~Pi, Y.-l. Zhang, Q.-G. Huang and M.~Sasaki, \emph{{Scalaron from
  $R^2$-gravity as a heavy field}},
  \href{http://dx.doi.org/10.1088/1475-7516/2018/05/042}{\emph{JCAP} {\bf 05}
  (2018) 042}, [\href{http://arxiv.org/abs/1712.09896}{{\tt 1712.09896}}].

\bibitem{Garcia-Bellido:1996mdl}
J.~Garcia-Bellido, A.~D. Linde and D.~Wands, \emph{{Density perturbations and
  black hole formation in hybrid inflation}},
  \href{http://dx.doi.org/10.1103/PhysRevD.54.6040}{\emph{Phys. Rev. D} {\bf
  54} (1996) 6040--6058}, [\href{http://arxiv.org/abs/astro-ph/9605094}{{\tt
  astro-ph/9605094}}].

\bibitem{Frampton:2010sw}
P.~H. Frampton, M.~Kawasaki, F.~Takahashi and T.~T. Yanagida, \emph{{Primordial
  Black Holes as All Dark Matter}},
  \href{http://dx.doi.org/10.1088/1475-7516/2010/04/023}{\emph{JCAP} {\bf 04}
  (2010) 023}, [\href{http://arxiv.org/abs/1001.2308}{{\tt 1001.2308}}].

\bibitem{Palma:2020ejf}
G.~A. Palma, S.~Sypsas and C.~Zenteno, \emph{{Seeding primordial black holes in
  multifield inflation}},
  \href{http://dx.doi.org/10.1103/PhysRevLett.125.121301}{\emph{Phys. Rev.
  Lett.} {\bf 125} (2020) 121301}, [\href{http://arxiv.org/abs/2004.06106}{{\tt
  2004.06106}}].

\bibitem{Fumagalli:2020adf}
J.~Fumagalli, S.~Renaux-Petel, J.~W. Ronayne and L.~T. Witkowski,
  \emph{{Turning in the landscape: a new mechanism for generating Primordial
  Black Holes}},  \href{http://arxiv.org/abs/2004.08369}{{\tt 2004.08369}}.

\bibitem{Braglia:2020eai}
M.~Braglia, D.~K. Hazra, F.~Finelli, G.~F. Smoot, L.~Sriramkumar and A.~A.
  Starobinsky, \emph{{Generating PBHs and small-scale GWs in two-field models
  of inflation}},
  \href{http://dx.doi.org/10.1088/1475-7516/2020/08/001}{\emph{JCAP} {\bf 08}
  (2020) 001}, [\href{http://arxiv.org/abs/2005.02895}{{\tt 2005.02895}}].

\bibitem{Chen:2019zza}
C.~Chen and Y.-F. Cai, \emph{{Primordial black holes from sound speed resonance
  in the inflaton-curvaton mixed scenario}},
  \href{http://dx.doi.org/10.1088/1475-7516/2019/10/068}{\emph{JCAP} {\bf 10}
  (2019) 068}, [\href{http://arxiv.org/abs/1908.03942}{{\tt 1908.03942}}].

\bibitem{Cai:2019jah}
Y.-F. Cai, C.~Chen, X.~Tong, D.-G. Wang and S.-F. Yan, \emph{{When Primordial
  Black Holes from Sound Speed Resonance Meet a Stochastic Background of
  Gravitational Waves}},
  \href{http://dx.doi.org/10.1103/PhysRevD.100.043518}{\emph{Phys. Rev. D} {\bf
  100} (2019) 043518}, [\href{http://arxiv.org/abs/1902.08187}{{\tt
  1902.08187}}].

\bibitem{Cai:2019bmk}
R.-G. Cai, Z.-K. Guo, J.~Liu, L.~Liu and X.-Y. Yang, \emph{{Primordial black
  holes and gravitational waves from parametric amplification of curvature
  perturbations}},
  \href{http://dx.doi.org/10.1088/1475-7516/2020/06/013}{\emph{JCAP} {\bf 06}
  (2020) 013}, [\href{http://arxiv.org/abs/1912.10437}{{\tt 1912.10437}}].

\bibitem{Chen:2020uhe}
C.~Chen, X.-H. Ma and Y.-F. Cai, \emph{{Dirac-Born-Infeld realization of sound
  speed resonance mechanism for primordial black holes}},
  \href{http://dx.doi.org/10.1103/PhysRevD.102.063526}{\emph{Phys. Rev. D} {\bf
  102} (2020) 063526}, [\href{http://arxiv.org/abs/2003.03821}{{\tt
  2003.03821}}].

\bibitem{Fumagalli:2020nvq}
J.~Fumagalli, S.~Renaux-Petel and L.~T. Witkowski, \emph{{Oscillations in the
  stochastic gravitational wave background from sharp features and particle
  production during inflation}},
  \href{http://dx.doi.org/10.1088/1475-7516/2021/08/030}{\emph{JCAP} {\bf 08}
  (2021) 030}, [\href{http://arxiv.org/abs/2012.02761}{{\tt 2012.02761}}].

\bibitem{Caprini:2015zlo}
C.~Caprini et~al., \emph{{Science with the space-based interferometer eLISA.
  II: Gravitational waves from cosmological phase transitions}},
  \href{http://dx.doi.org/10.1088/1475-7516/2016/04/001}{\emph{JCAP} {\bf 04}
  (2016) 001}, [\href{http://arxiv.org/abs/1512.06239}{{\tt 1512.06239}}].

\bibitem{Auclair:2019wcv}
P.~Auclair et~al., \emph{{Probing the gravitational wave background from cosmic
  strings with LISA}},
  \href{http://dx.doi.org/10.1088/1475-7516/2020/04/034}{\emph{JCAP} {\bf 04}
  (2020) 034}, [\href{http://arxiv.org/abs/1909.00819}{{\tt 1909.00819}}].

\bibitem{Mazumdar:2018dfl}
A.~Mazumdar and G.~White, \emph{{Review of cosmic phase transitions: their
  significance and experimental signatures}},
  \href{http://dx.doi.org/10.1088/1361-6633/ab1f55}{\emph{Rept. Prog. Phys.}
  {\bf 82} (2019) 076901}, [\href{http://arxiv.org/abs/1811.01948}{{\tt
  1811.01948}}].

\bibitem{Garcia-Bellido:2017aan}
J.~Garcia-Bellido, M.~Peloso and C.~Unal, \emph{{Gravitational Wave signatures
  of inflationary models from Primordial Black Hole Dark Matter}},
  \href{http://dx.doi.org/10.1088/1475-7516/2017/09/013}{\emph{JCAP} {\bf 09}
  (2017) 013}, [\href{http://arxiv.org/abs/1707.02441}{{\tt 1707.02441}}].

\bibitem{Bartolo:2018evs}
N.~Bartolo, V.~De~Luca, G.~Franciolini, A.~Lewis, M.~Peloso and A.~Riotto,
  \emph{{Primordial Black Hole Dark Matter: LISA Serendipity}},
  \href{http://dx.doi.org/10.1103/PhysRevLett.122.211301}{\emph{Phys. Rev.
  Lett.} {\bf 122} (2019) 211301}, [\href{http://arxiv.org/abs/1810.12218}{{\tt
  1810.12218}}].

\bibitem{Bartolo:2018rku}
N.~Bartolo, V.~De~Luca, G.~Franciolini, M.~Peloso, D.~Racco and A.~Riotto,
  \emph{{Testing primordial black holes as dark matter with LISA}},
  \href{http://dx.doi.org/10.1103/PhysRevD.99.103521}{\emph{Phys. Rev. D} {\bf
  99} (2019) 103521}, [\href{http://arxiv.org/abs/1810.12224}{{\tt
  1810.12224}}].

\bibitem{bradley_j_kavanagh_2019_3538999}
B.~J. Kavanagh, \emph{bradkav/pbhbounds: Release version},  Nov., 2019.
\newblock 10.5281/zenodo.3538999.

\bibitem{Byrnes:2018clq}
C.~T. Byrnes, M.~Hindmarsh, S.~Young and M.~R.~S. Hawkins, \emph{{Primordial
  black holes with an accurate QCD equation of state}},
  \href{http://dx.doi.org/10.1088/1475-7516/2018/08/041}{\emph{JCAP} {\bf 08}
  (2018) 041}, [\href{http://arxiv.org/abs/1801.06138}{{\tt 1801.06138}}].

\bibitem{Gow:2020bzo}
A.~D. Gow, C.~T. Byrnes, P.~S. Cole and S.~Young, \emph{{The power spectrum on
  small scales: Robust constraints and comparing PBH methodologies}},
  \href{http://dx.doi.org/10.1088/1475-7516/2021/02/002}{\emph{JCAP} {\bf 02}
  (2021) 002}, [\href{http://arxiv.org/abs/2008.03289}{{\tt 2008.03289}}].

\bibitem{Carr:2017jsz}
B.~Carr, M.~Raidal, T.~Tenkanen, V.~Vaskonen and H.~Veerm\"ae,
  \emph{{Primordial black hole constraints for extended mass functions}},
  \href{http://dx.doi.org/10.1103/PhysRevD.96.023514}{\emph{Phys. Rev. D} {\bf
  96} (2017) 023514}, [\href{http://arxiv.org/abs/1705.05567}{{\tt
  1705.05567}}].

\bibitem{Smith:2019wny}
T.~L. Smith and R.~Caldwell, \emph{{LISA for Cosmologists: Calculating the
  Signal-to-Noise Ratio for Stochastic and Deterministic Sources}},
  \href{http://dx.doi.org/10.1103/PhysRevD.100.104055}{\emph{Phys. Rev. D} {\bf
  100} (2019) 104055}, [\href{http://arxiv.org/abs/1908.00546}{{\tt
  1908.00546}}].

\bibitem{Lewicki:2021kmu}
M.~Lewicki and V.~Vaskonen, \emph{{Impact of LIGO-Virgo binaries on
  gravitational wave background searches}},
  \href{http://arxiv.org/abs/2111.05847}{{\tt 2111.05847}}.

\bibitem{Smyth:2019whb}
N.~Smyth, S.~Profumo, S.~English, T.~Jeltema, K.~McKinnon and P.~Guhathakurta,
  \emph{{Updated Constraints on Asteroid-Mass Primordial Black Holes as Dark
  Matter}}, \href{http://dx.doi.org/10.1103/PhysRevD.101.063005}{\emph{Phys.
  Rev. D} {\bf 101} (2020) 063005},
  [\href{http://arxiv.org/abs/1910.01285}{{\tt 1910.01285}}].

\bibitem{EROS-2:2006ryy}
{\scshape EROS-2} collaboration, P.~Tisserand et~al., \emph{{Limits on the
  Macho Content of the Galactic Halo from the EROS-2 Survey of the Magellanic
  Clouds}}, \href{http://dx.doi.org/10.1051/0004-6361:20066017}{\emph{Astron.
  Astrophys.} {\bf 469} (2007) 387--404},
  [\href{http://arxiv.org/abs/astro-ph/0607207}{{\tt astro-ph/0607207}}].

\bibitem{Croon:2020ouk}
D.~Croon, D.~McKeen, N.~Raj and Z.~Wang, \emph{{Subaru-HSC through a different
  lens: Microlensing by extended dark matter structures}},
  \href{http://dx.doi.org/10.1103/PhysRevD.102.083021}{\emph{Phys. Rev. D} {\bf
  102} (2020) 083021}, [\href{http://arxiv.org/abs/2007.12697}{{\tt
  2007.12697}}].

\bibitem{Griest:2013aaa}
K.~Griest, A.~M. Cieplak and M.~J. Lehner, \emph{{Experimental Limits on
  Primordial Black Hole Dark Matter from the First 2 yr of Kepler Data}},
  \href{http://dx.doi.org/10.1088/0004-637X/786/2/158}{\emph{Astrophys. J.}
  {\bf 786} (2014) 158}, [\href{http://arxiv.org/abs/1307.5798}{{\tt
  1307.5798}}].

\bibitem{Oguri:2017ock}
M.~Oguri, J.~M. Diego, N.~Kaiser, P.~L. Kelly and T.~Broadhurst,
  \emph{{Understanding caustic crossings in giant arcs: characteristic scales,
  event rates, and constraints on compact dark matter}},
  \href{http://dx.doi.org/10.1103/PhysRevD.97.023518}{\emph{Phys. Rev. D} {\bf
  97} (2018) 023518}, [\href{http://arxiv.org/abs/1710.00148}{{\tt
  1710.00148}}].

\bibitem{Niikura:2019kqi}
H.~Niikura, M.~Takada, S.~Yokoyama, T.~Sumi and S.~Masaki, \emph{{Constraints
  on Earth-mass primordial black holes from OGLE 5-year microlensing events}},
  \href{http://dx.doi.org/10.1103/PhysRevD.99.083503}{\emph{Phys. Rev. D} {\bf
  99} (2019) 083503}, [\href{http://arxiv.org/abs/1901.07120}{{\tt
  1901.07120}}].

\bibitem{Macho:2000nvd}
{\scshape Macho} collaboration, R.~A. Allsman et~al., \emph{{MACHO project
  limits on black hole dark matter in the 1-30 solar mass range}},
  \href{http://dx.doi.org/10.1086/319636}{\emph{Astrophys. J. Lett.} {\bf 550}
  (2001) L169}, [\href{http://arxiv.org/abs/astro-ph/0011506}{{\tt
  astro-ph/0011506}}].

\bibitem{Manshanden:2018tze}
J.~Manshanden, D.~Gaggero, G.~Bertone, R.~M.~T. Connors and M.~Ricotti,
  \emph{{Multi-wavelength astronomical searches for primordial black holes}},
  \href{http://dx.doi.org/10.1088/1475-7516/2019/06/026}{\emph{JCAP} {\bf 06}
  (2019) 026}, [\href{http://arxiv.org/abs/1812.07967}{{\tt 1812.07967}}].

\bibitem{Lu:2020bmd}
P.~Lu, V.~Takhistov, G.~B. Gelmini, K.~Hayashi, Y.~Inoue and A.~Kusenko,
  \emph{{Constraining Primordial Black Holes with Dwarf Galaxy Heating}},
  \href{http://dx.doi.org/10.3847/2041-8213/abdcb6}{\emph{Astrophys. J. Lett.}
  {\bf 908} (2021) L23}, [\href{http://arxiv.org/abs/2007.02213}{{\tt
  2007.02213}}].

\bibitem{Serpico:2020ehh}
P.~D. Serpico, V.~Poulin, D.~Inman and K.~Kohri, \emph{{Cosmic microwave
  background bounds on primordial black holes including dark matter halo
  accretion}},
  \href{http://dx.doi.org/10.1103/PhysRevResearch.2.023204}{\emph{Phys. Rev.
  Res.} {\bf 2} (2020) 023204}, [\href{http://arxiv.org/abs/2002.10771}{{\tt
  2002.10771}}].

\bibitem{Hektor:2018qqw}
A.~Hektor, G.~H\"utsi, L.~Marzola, M.~Raidal, V.~Vaskonen and H.~Veerm\"ae,
  \emph{{Constraining Primordial Black Holes with the EDGES 21-cm Absorption
  Signal}}, \href{http://dx.doi.org/10.1103/PhysRevD.98.023503}{\emph{Phys.
  Rev. D} {\bf 98} (2018) 023503}, [\href{http://arxiv.org/abs/1803.09697}{{\tt
  1803.09697}}].

\bibitem{Hutsi:2020sol}
G.~H\"utsi, M.~Raidal, V.~Vaskonen and H.~Veerm\"ae, \emph{{Two populations of
  LIGO-Virgo black holes}},
  \href{http://dx.doi.org/10.1088/1475-7516/2021/03/068}{\emph{JCAP} {\bf 03}
  (2021) 068}, [\href{http://arxiv.org/abs/2012.02786}{{\tt 2012.02786}}].

\bibitem{Carr:2009jm}
B.~J. Carr, K.~Kohri, Y.~Sendouda and J.~Yokoyama, \emph{{New cosmological
  constraints on primordial black holes}},
  \href{http://dx.doi.org/10.1103/PhysRevD.81.104019}{\emph{Phys. Rev. D} {\bf
  81} (2010) 104019}, [\href{http://arxiv.org/abs/0912.5297}{{\tt 0912.5297}}].

\bibitem{Clark:2016nst}
S.~Clark, B.~Dutta, Y.~Gao, L.~E. Strigari and S.~Watson, \emph{{Planck
  Constraint on Relic Primordial Black Holes}},
  \href{http://dx.doi.org/10.1103/PhysRevD.95.083006}{\emph{Phys. Rev. D} {\bf
  95} (2017) 083006}, [\href{http://arxiv.org/abs/1612.07738}{{\tt
  1612.07738}}].

\bibitem{Boudaud:2018hqb}
M.~Boudaud and M.~Cirelli, \emph{{Voyager 1 $e^\pm$ Further Constrain
  Primordial Black Holes as Dark Matter}},
  \href{http://dx.doi.org/10.1103/PhysRevLett.122.041104}{\emph{Phys. Rev.
  Lett.} {\bf 122} (2019) 041104}, [\href{http://arxiv.org/abs/1807.03075}{{\tt
  1807.03075}}].

\bibitem{Clark:2018ghm}
S.~Clark, B.~Dutta, Y.~Gao, Y.-Z. Ma and L.~E. Strigari, \emph{{21 cm limits on
  decaying dark matter and primordial black holes}},
  \href{http://dx.doi.org/10.1103/PhysRevD.98.043006}{\emph{Phys. Rev. D} {\bf
  98} (2018) 043006}, [\href{http://arxiv.org/abs/1803.09390}{{\tt
  1803.09390}}].

\bibitem{Dasgupta:2019cae}
B.~Dasgupta, R.~Laha and A.~Ray, \emph{{Neutrino and positron constraints on
  spinning primordial black hole dark matter}},
  \href{http://dx.doi.org/10.1103/PhysRevLett.125.101101}{\emph{Phys. Rev.
  Lett.} {\bf 125} (2020) 101101}, [\href{http://arxiv.org/abs/1912.01014}{{\tt
  1912.01014}}].

\bibitem{DeRocco:2019fjq}
W.~DeRocco and P.~W. Graham, \emph{{Constraining Primordial Black Hole
  Abundance with the Galactic 511 keV Line}},
  \href{http://dx.doi.org/10.1103/PhysRevLett.123.251102}{\emph{Phys. Rev.
  Lett.} {\bf 123} (2019) 251102}, [\href{http://arxiv.org/abs/1906.07740}{{\tt
  1906.07740}}].

\bibitem{Laha:2019ssq}
R.~Laha, \emph{{Primordial Black Holes as a Dark Matter Candidate Are Severely
  Constrained by the Galactic Center 511 keV $\gamma$ -Ray Line}},
  \href{http://dx.doi.org/10.1103/PhysRevLett.123.251101}{\emph{Phys. Rev.
  Lett.} {\bf 123} (2019) 251101}, [\href{http://arxiv.org/abs/1906.09994}{{\tt
  1906.09994}}].

\bibitem{Laha:2020ivk}
R.~Laha, J.~B. Mu\~noz and T.~R. Slatyer, \emph{{INTEGRAL constraints on
  primordial black holes and particle dark matter}},
  \href{http://dx.doi.org/10.1103/PhysRevD.101.123514}{\emph{Phys. Rev. D} {\bf
  101} (2020) 123514}, [\href{http://arxiv.org/abs/2004.00627}{{\tt
  2004.00627}}].

\bibitem{Laha:2020vhg}
R.~Laha, P.~Lu and V.~Takhistov, \emph{{Gas heating from spinning and
  non-spinning evaporating primordial black holes}},
  \href{http://dx.doi.org/10.1016/j.physletb.2021.136459}{\emph{Phys. Lett. B}
  {\bf 820} (2021) 136459}, [\href{http://arxiv.org/abs/2009.11837}{{\tt
  2009.11837}}].

\bibitem{Saha:2021pqf}
A.~K. Saha and R.~Laha, \emph{{Sensitivities on non-spinning and spinning
  primordial black hole dark matter with global 21 cm troughs}},
  \href{http://arxiv.org/abs/2112.10794}{{\tt 2112.10794}}.

\bibitem{Mittal:2021egv}
S.~Mittal, A.~Ray, G.~Kulkarni and B.~Dasgupta, \emph{{Constraining primordial
  black holes as dark matter using the global 21-cm signal with X-ray heating
  and excess radio background}},
  \href{http://dx.doi.org/10.1088/1475-7516/2022/03/030}{\emph{JCAP} {\bf 03}
  (2022) 030}, [\href{http://arxiv.org/abs/2107.02190}{{\tt 2107.02190}}].

\bibitem{Monroy:2014}
M.~A. Monroy-Rodríguez and C.~Allen, \emph{The end of the macho era,
  revisited: New limits on macho masses from halo wide binaries},
  \href{http://dx.doi.org/10.1088/0004-637x/790/2/159}{\emph{The Astrophysical
  Journal} {\bf 790} (Jul, 2014) 159},
  [\href{http://arxiv.org/abs/1406.5169}{{\tt 1406.5169}}].

\bibitem{Brandt:2016aco}
T.~D. Brandt, \emph{{Constraints on MACHO Dark Matter from Compact Stellar
  Systems in Ultra-Faint Dwarf Galaxies}},
  \href{http://dx.doi.org/10.3847/2041-8205/824/2/L31}{\emph{Astrophys. J.
  Lett.} {\bf 824} (2016) L31}, [\href{http://arxiv.org/abs/1605.03665}{{\tt
  1605.03665}}].

\bibitem{Lewis:1999bs}
A.~Lewis, A.~Challinor and A.~Lasenby, \emph{{Efficient computation of CMB
  anisotropies in closed FRW models}},
  \href{http://dx.doi.org/10.1086/309179}{\emph{Astrophys. J.} {\bf 538} (2000)
  473--476}, [\href{http://arxiv.org/abs/astro-ph/9911177}{{\tt
  astro-ph/9911177}}].

\bibitem{Kudoh:2004he}
H.~Kudoh and A.~Taruya, \emph{{Probing anisotropies of gravitational-wave
  backgrounds with a space-based interferometer: Geometric properties of
  antenna patterns and their angular power}},
  \href{http://dx.doi.org/10.1103/PhysRevD.71.024025}{\emph{Phys. Rev. D} {\bf
  71} (2005) 024025}, [\href{http://arxiv.org/abs/gr-qc/0411017}{{\tt
  gr-qc/0411017}}].

\bibitem{Alonso:2020rar}
D.~Alonso, C.~R. Contaldi, G.~Cusin, P.~G. Ferreira and A.~I. Renzini,
  \emph{{Noise angular power spectrum of gravitational wave background
  experiments}},
  \href{http://dx.doi.org/10.1103/PhysRevD.101.124048}{\emph{Phys. Rev. D} {\bf
  101} (2020) 124048}, [\href{http://arxiv.org/abs/2005.03001}{{\tt
  2005.03001}}].

\bibitem{Contaldi:2020rht}
C.~R. Contaldi, M.~Pieroni, A.~I. Renzini, G.~Cusin, N.~Karnesis, M.~Peloso
  et~al., \emph{{Maximum likelihood map-making with the Laser Interferometer
  Space Antenna}},
  \href{http://dx.doi.org/10.1103/PhysRevD.102.043502}{\emph{Phys. Rev. D} {\bf
  102} (2020) 043502}, [\href{http://arxiv.org/abs/2006.03313}{{\tt
  2006.03313}}].

\bibitem{Cusin:2022cbb}
G.~Cusin and G.~Tasinato, \emph{{Doppler boosting the stochastic gravitational
  wave background}},  \href{http://arxiv.org/abs/2201.10464}{{\tt 2201.10464}}.

\bibitem{Bartolo:2022pez}
N.~Bartolo et~al., \emph{{Probing Anisotropies of the Stochastic Gravitational
  Wave Background with LISA}},  \href{http://arxiv.org/abs/2201.08782}{{\tt
  2201.08782}}.

\bibitem{Ruan:2020smc}
W.-H. Ruan, C.~Liu, Z.-K. Guo, Y.-L. Wu and R.-G. Cai, \emph{{The LISA-Taiji
  network}}, \href{http://dx.doi.org/10.1038/s41550-019-1008-4}{\emph{Nature
  Astron.} {\bf 4} (2020) 108--109},
  [\href{http://arxiv.org/abs/2002.03603}{{\tt 2002.03603}}].

\bibitem{Smith:2016jqs}
T.~L. Smith and R.~Caldwell, \emph{{Sensitivity to a Frequency-Dependent
  Circular Polarization in an Isotropic Stochastic Gravitational Wave
  Background}}, \href{http://dx.doi.org/10.1103/PhysRevD.95.044036}{\emph{Phys.
  Rev. D} {\bf 95} (2017) 044036}, [\href{http://arxiv.org/abs/1609.05901}{{\tt
  1609.05901}}].

\bibitem{Tegmark:1997vs}
M.~Tegmark, \emph{{CMB mapping experiments: A Designer's guide}},
  \href{http://dx.doi.org/10.1103/PhysRevD.56.4514}{\emph{Phys. Rev. D} {\bf
  56} (1997) 4514--4529}, [\href{http://arxiv.org/abs/astro-ph/9705188}{{\tt
  astro-ph/9705188}}].

\bibitem{Tegmark:1999ke}
M.~Tegmark, D.~J. Eisenstein, W.~Hu and A.~de~Oliveira-Costa,
  \emph{{Foregrounds and forecasts for the cosmic microwave background}},
  \href{http://dx.doi.org/10.1086/308348}{\emph{Astrophys. J.} {\bf 530} (2000)
  133--165}, [\href{http://arxiv.org/abs/astro-ph/9905257}{{\tt
  astro-ph/9905257}}].

\bibitem{Mukhanov:2005sc}
V.~Mukhanov, \emph{{Physical Foundations of Cosmology}}.
\newblock Cambridge University Press, Oxford, 2005.

\bibitem{Wald:1984rg}
R.~M. Wald, \emph{{General Relativity}}.
\newblock Chicago Univ. Pr., Chicago, USA, 1984.
\newblock 10.7208/chicago/9780226870373.001.0001.

\bibitem{Alba:2017clw}
V.~Alba, \emph{{Aspects of scale invariance in Physics and Biology}}.
\newblock PhD thesis, Princeton U., 9, 2017.

\end{thebibliography}\endgroup

\end{document}